\documentclass[twocolumn,11pt]{aastex61}

\usepackage{mathtools} % includes amsmath
\usepackage{natbib}
\usepackage{color}
\usepackage{enumitem}
\usepackage{algorithm,algpseudocode}
\usepackage{bbold}
\usepackage{gensymb}
\usepackage{float} % for more control on location of figures
\usepackage{xspace}

\graphicspath{{figs/}}

\newcommand{\Lphi}[1]{\mathcal L(#1)}
\newcommand{\Ccmb}{\mathcal C_f}
\newcommand{\preCcmbpost}[2]{ #1{\mathcal C}_{f,#2} }
\newcommand{\preCcmb}[1]{ #1{\mathcal C}_f }
\newcommand{\Ccmbtheta}{\mathcal C_f(\theta)}

\newcommand{\Cn}{\mathcal C_{n}}

\newcommand{\Clen}{\mathcal C_{\phi}}
\newcommand{\Clentheta}{\mathcal C_\phi(\theta)}

\newcommand{\logdet}{\log\det\xspace}
\DeclareMathOperator*{\argmax}{arg\,max} 

\begin{document}

\title{Bayesian delensing of CMB temperature and polarization}

\author{Marius Millea} 
\affiliation{Institut d’Astrophysique de Paris (IAP), UMR 7095, CNRS – UPMC Université Paris 6, Sorbonne Universités,
98bis boulevard Arago, F-75014 Paris, France}
\affiliation{Institut Lagrange de Paris (ILP), Sorbonne Universités,
98bis boulevard Arago, F-75014 Paris, France}

\author{Ethan Anderes}
\affiliation{Department of Statistics, University of California, Davis, CA 95616, USA}

\author{Benjamin D. Wandelt}
\affiliation{Institut d’Astrophysique de Paris (IAP), UMR 7095, CNRS – UPMC Université Paris 6, Sorbonne Universités,
98bis boulevard Arago, F-75014 Paris, France}
\affiliation{Institut Lagrange de Paris (ILP), Sorbonne Universités,
98bis boulevard Arago, F-75014 Paris, France}
\affiliation{Department of Physics and Astronomy, University of Illinois at Urbana-Champaign, 1002 W Green St, Urbana, IL 61801, USA}
\affiliation{Center for Computational Astrophysics, Flatiron Institute, 162 5th Avenue, 10010, New York, NY, USA}

\correspondingauthor{Marius Millea}
\email{mariusmillea@gmail.com}

\begin{abstract}

We develop the first algorithm able to jointly compute the maximum {\it a
posteriori} estimate of the Cosmic Microwave Background (CMB) temperature and
polarization fields, the gravitational potential by which they are lensed, and
cosmological parameters such as the tensor-to-scalar ratio, $r$. This is an
important step towards sampling from the joint posterior probability function of
these quantities, which, assuming Gaussianity of the CMB fields and lensing
potential, contains all available cosmological information and would yield
theoretically optimal constraints. Attaining such optimal constraints will be
crucial for next-generation CMB surveys like CMB-S4, where limits on $r$ could
be improved by factors of a few over currently used sub-optimal quadratic
estimators. The maximization procedure described here depends on a newly
developed lensing algorithm, which we term \textsc{LenseFlow}, and which lenses
a map by solving a system of ordinary differential equations. This description
has conceptual advantages, such as allowing us to give a simple non-perturbative
proof that the lensing determinant is equal to unity in the weak-lensing regime.
The algorithm itself maintains this property even on pixelized maps, which is
crucial for our purposes and unique to \textsc{LenseFlow} as compared to other
lensing algorithms we have tested. It also has other useful properties such as
that it can be trivially inverted (i.e. delensing) for the same computational
cost as the forward operation, and can be used to compute lensing adjoint,
Jacobian, and Hessian operators. We test and validate the maximization procedure
on flat-sky simulations covering up to 600\,deg$^2$ with non-uniform noise and
masking.

\end{abstract}


\section{Introduction} \label{sec:intro}

Weak gravitational lensing of the Cosmic Microwave Background (CMB) by
intervening large scale structure plays and will continue to play a crucial role
in the ability of cosmological observations to constrain fundamental physics.
For example, the gravitational lensing effect already allows a completely
independent confirmation of the existence of dark energy from the CMB alone
\citep{sherwin2011}, and future experiments such as CMB-S4 are predicted to map
out the gravitational lensing potential field, $\phi$, precisely enough to
measure for the first time the absolute neutrino mass scale and potentially
differentiate the two possible mass hierarchies \citep{abazajian2013}. A wealth
of cosmological and astrophysical information can also be extracted from these
lensing potential maps in cross-correlation with other datasets
\citep[see e.g.][]{abazajian2016}.

The most profound impact from CMB lensing on our understanding of the universe,
however, may come not from measuring the effect, per se, but rather from our
ability to remove it. Lensing aliases $E$-mode polarization into $B$-modes,
which can obscure the primordial $B$ signal expected to come from gravitational
waves produced during inflation. Due to its unique signature, it is possible to
undo the lensing effect, a process usually called ``delensing''. This will be
crucial to placing the tightest possible constraints on the amplitude, $r$, of
the gravitational wave $B$-modes. If detected, the primordial signal would offer
an unprecedented window into the extremely early universe and to energy scales
impossible to probe with terrestrial particle accelerators.

Delensing of both $T$ and $E$ can also be useful as it leads to a sharpening of
the acoustic peaks. This in turn makes it easier to measure their phase and
could lead to detecting or ruling out the presence of extra species of
relativistic particles in the universe \citep{green2016}.

Despite the important role delensing is expected to play in future CMB
constraints, currently no workable fully optimal delensing algorithm exists. To
date, all delensing analyses on real data have been based on a quadratic
estimate of the lensing potential \citep{hu2002,okamoto2003}. While the
quadratic estimator is nearly optimal at current noise levels, it will become
significantly sub-optimal once noise levels cross below the $\sim5\mu$K-arcmin
effective noise level of the lensing contribution (exactly when delensing
becomes most important). The sub-optimality of the quadratic estimate stems from
the fact that the total $B$-mode power is a source of noise for the estimator,
meaning the results can be improved by repeatedly using the lensing potential
estimate to delense the data and then re-estimating the lensing potential. Such
iterative delensing algorithms have been discussed in some form in e.g.
\cite{kesden2002, knox2002, hanson2010, smith2012}.

Two concrete iterative delensing examples which can be considered precursors to
our work have been given by \cite{hirata2003} and \cite{carron2017}. In a
similar manner to iterating a quadratic estimate, both of these algorithms
iteratively maximize the Bayesian posterior probability
$\mathcal{P}(\phi\,|\,d,r)$, where $\phi$ is the lensing potential and $d$ is
the CMB temperature and polarization data.\footnote{These algorithms actually
produce estimates of the full lensing displacement vector field, not just of
$\phi$ which gives only the curl-free part in the Helmholtz decomposition of the
displacement. For simplicity, we will ignore the divergence-free component
throughout this work as it is expected to be too small to significantly impact
the $\phi$ reconstruction at CMB-S4 noise levels \citep{hirata2003}, but it is
straight-forward to introduce it in our equations alongside $\phi$.} In terms of
the end product, the two differ largely in that the latter algorithm computes
the exact maximum and was demonstrated to be robust even in the presence of
masking. These works greatly improve the optimality of the lensing
reconstruction and represent key advances in CMB lensing analysis. However,
neither estimate is truly optimal in the least-squared sense, and neither
readily produces an estimate of an unlensed map nor of $r$. Indeed, since the
temperature and polarization fields themselves are implicitly marginalized over
in $\mathcal{P}(\phi\,|\,d,r)$, unlensed fields are not estimated at all by
these procedures. The resulting best-fit $\phi$ could be used to delense the
data, but as we will discuss, this resulting delensed data does not have any
Bayesian interpretation. The delensed map could be taken as an estimator, but
would still require simulations to debias and quantitify uncertainty, similarly
as for the quadratic estimate but with a more costly procedure to simulate. More
importantly, it is not entirely clear how to do this at all because these
simulations would depend on $r$, the quantity we are trying to estimate in the
first place. Indeed, in their stated form both algorithms take $r$ as given,
rather than jointly estimating it or marginalizing over it.

A conceptually straightforward solution to these issues which would yield
optimal constraints on all of these quantities is to obtain samples from the
{\it joint} Bayesian posterior probability function,
$\mathcal{P}(f,\phi,r\,|\,d)$, including both the unlensed fields,
$f\equiv(T,Q,U)$, and the tensor-to-scalar ratio, $r$. Here, we present the
first algorithm which is able to efficiently {\it maximize} this probability
distribution, an important advancement towards the ultimate goal of obtaining
samples. Additionally, the best-fit computed here can be used as an
initialization for a sampler, and we expect that a good starting point will be
important due to the high dimensionally of the problem (the number of dimensions
here being the number of map pixels, which can be in the millions). Although we
do not expect joint sampling to be without challenges, it has already been
demonstrated on temperature-only data by \cite{anderes2015}, and we view the
techniques developed here as having solved the more difficult aspects of the
problem of extending to polarization. We leave full discussion of sampling with
temperature and polarization for a follow-up work, here discussing mainly
maximization.

The results here also differ from \cite{anderes2015} by exploring $r$ as a free
parameter. In some sense it is quite easy to maximize over $r$, since we can
trivially parallelize the maximization over $f$ and $\phi$ across a grid of $r$
values. Doing so, we will show that maximum a-posteriori (MAP) estimate of $r$
in this joint case does not have good properties as an estimator. We will thus
focus most of our discussion on $\mathcal{P}(f,\phi\,|\,d,r)$.

As opposed to exact maximization of $\mathcal P(\phi\,|\,d,r)$ which was solved
by \cite{carron2017}, maximization of $\mathcal P(f,\phi\,|\,d,r)$ is more
difficult not just because of the increased dimensionality of the problem, but
because $f$ is highly correlated with $\phi$. Intuitively, this is simply
because an observed hot-spot at some position could be a true hot-spot there
with no lensing, or a nearby hot-spot deflected to that position by lensing.
This degeneracy leads to extremely slow convergence unless the correlations are
carefully taken into account. We find an advantageous way to do so is to
reparametrize the posterior probability function in terms of the lensed fields
(denoted by $\tilde f$) instead of the of the unlensed ones, similarly as in
\cite{anderes2015}. This greatly reduces the correlations, but the change of
variables introduces a term in the posterior probability which depends on the
determinant of the lensing operator. Having to calculate this quantity might
render the reparameterization ultimately useless in practice. However, we are
able to develop a new and accurate pixelized lensing approximation which we call
\textsc{LenseFlow} which is area-preserving, i.e. for which the determinant is
unity and can thus be ignored.

We use this in a maximization algorithm that can be regarded as an approximate
coordinate descent, meaning we alternate updating $\tilde f$ with $\phi$ held
constant then updating $\phi$ with the $\tilde f$ held constant. The former step
amounts to a straight-forward Wiener filter, and the latter step can be
approximated with a quasi Newton-Raphson step. As we will show, a fundamental
advantage of the lensed parametrization (in addition to reducing correlations),
is that it removes all explicit dependence on data or instrument from this
latter step. These two steps are repeated until convergence to the exact joint
posterior maximum, which, depending on the exact data configuration and
complexity of masking, we can achieve in 30 minutes to tens of hours on a single
multi-core CPU for maps as large as $\sim$600\,deg$^2$ (with 3 arcmin pixels).

By contrast, the maximization procedure described in \cite{carron2017} requires
orders of magnitude more computation time due to the costly calculation of a
determinant gradient term. We will discuss why our seemingly complicating
addition of jointly estimating $f$ actually makes the problem computationally
easier, and what the trade-off has been in not computing this determinant.
Furthermore, we will argue that even if one was only interested in posterior
samples of $r$, it will still be computationally simpler to obtain them by
sampling the joint posterior rather than the one marginalized over $f$.

The maximization makes use of exact posterior gradients, which are computable
with \textsc{LenseFlow}. We show that even though Hessians of the posterior can
not be stored in practice, their action on vectors can be efficiently
calculated, a fact which is perhaps not widely appreciated. Although we do not
use them here, Hessians could be quite beneficial to sampling algorithms.

Our code is available
publicly.\footnote{\url{https://www.github.com/marius311/CMBLensing.jl}} It is
written in the \textsc{Julia} programming language \citep{bezanson2017}, making
it fast while maintaining flexibility and readability. The link also contains a
\textsc{Jupyter} notebook with a 128$\times$128 pixel maximization example which
completes in around two minutes on a modern laptop.

We begin the paper by deriving the joint Bayesian posterior in
Sec.~\ref{sec:posterior} and discussing how it is related to the marginalized
posterior in Sec.~\ref{sec:margnalized_posterior}. We then derive the coordinate
descent equations for the joint posterior maximization in
Sec.~\ref{sec:maximization}. We develop \textsc{LenseFlow}, its gradients, as
well as the proof that its determinant is unity in Sec.~\ref{sec: lenseflow}. We
show results on simulated data in Sec.~\ref{sec:results}. The results are broken
into several parts for clarity of presentation, first with only Fourier-space
masking in Sec.~\ref{sec:results_nomasking}, next with map-level masking as well
in Sec.~\ref{sec:results_masking}, and then with $r$ included as a free
parameter in Sec.~\ref{sec:results_r}. Finally, we revisit the discussion of
lensing determinant in more detail in Sec.~\ref{sec:lensdet} before concluding.

\section{The joint posterior probability}
\label{sec:posterior}

To start, we derive the target probability function that we seek to maximize in
this work, mainly the joint posterior probability of the unlensed CMB, the CMB
gravitational lensing potential, and the cosmological parameters.

Briefly summarizing our notation, we use $\phi$ for the gravitational lensing
potential and $f$ to describe a CMB field such as the temperature, $T$, or a
tuple including polarization Stokes parameters, such as $(Q,U)$ or $(T,Q,U)$.
Lensed fields are denoted with a tilde, $\tilde f$. Quantities like $\tilde f$,
$f$, or $\phi$ should be thought of as abstract vectors, meaning they can be
added and scaled without need to reference the basis in which they are
represented. Indeed, most of our equations are written without reference to
basis; at the few points where it {\it is} necessary to do so, we use $f(x)$ or
$f(l)$ to refer to the real-space or Fourier basis. We use the notation
$f^\dagger g$ to denote the inner product between fields $f$ and $g$, which is
defined to be a sum over products of corresponding temperature and polarization
pixels in $f$ and $g$. Linear operators on this resulting Hilbert space will be
capital letters, e.g. $\mathcal L$, and adjoint operators, $\mathcal L^\dagger$,
are defined as usual by the property that $f^\dagger (\mathcal L g) = (\mathcal
L^\dagger f)^\dagger g$ for all $f$ and $g$. We often use
$\mathcal{L}^{-\dagger}$ as shorthand for the inverse then adjoint of the
operator.

We model the data, $d$, as related to the true unlensed field, $f$, by a lensing
operation, $\Lphi{\phi} $, which is a linear operator dependent on the lensing
potential, and a noise contribution, $n$. Without loss of generality, we
implicitly absorb the deconvolution of any beam or instrumental transfer
function contribution into $n$; for real data analysis, these can in practice be
handled in whichever way is convenient. Thus we have,
\begin{align}
    d &= \Lphi{\phi} f + n \\
      &= \tilde f + n
\end{align}
Assuming the noise is a Gaussian random field with covariance $\Cn$, the
likelihood of the data is, up to an irrelevant normalization constant,
\begin{align}
    -2\log \mathcal P(d\,|\,f,\phi) = [d-\Lphi{\phi} f]^\dagger \Cn^{-1} [d-\Lphi{\phi} f]
\end{align}

By Bayes theorem, the posterior probability of $f$, $\phi$, and of any
cosmological parameters, $\theta$, is proportional to this likelihood times a
prior $\mathcal P(f,\phi,\theta)$,
\begin{align}
-2&\log \mathcal P(f,\phi,\theta \,|\, d)= \nonumber\\
&= -2\log \mathcal P(d\,|\,f,\phi) -2 \log \mathcal P(f,\phi,\theta) \nonumber\\
&= \big[d- \Lphi{\phi} f\big]^\dagger \Cn^{-1} \big[d-\Lphi{\phi} f\big]  \label{eq:lnP} \\
&\qquad\quad + f^\dagger \Ccmbtheta^{-1}\, f +  \phi^\dagger \Clentheta^{-1}\, \phi \nonumber\\
&\qquad\quad + \logdet \Ccmbtheta + \logdet \Clentheta  \nonumber
\end{align}

One is entirely free to chose the prior function to be as informative or
uninformative as desired, although {\it something} about $f$ must be specified
for a posterior constraint of $\phi$ to be produced. Here we adopt the prior
that both $f$ and $\phi$ are independent Gaussian random fields with covariance
given by $\Ccmb$ and $\Clen$, respectively, each of which may depend on some set
of cosmological parameters, $\theta$. We ignore any prior correlation between
$f$ and $\phi$, the most dominant expected contribution being at large scales in
temperature due to the late-time integrated Sachs-Wolfe effect. It is
straight-forward to include this in \eqref{eq:lnP}, but we have not done so for
simplicity and since it is likely too small to matter at the scales probed by
the patches of sky considered here. Additionally, as mentioned in
\cite{carron2017}, using a Gaussian prior on $\phi$ (and in our case, $f$) does
not outright erase from the reconstruction any non-Gaussianities that may be
present in $f$ and/or $\phi$ from various higher order effects. However, it does
mean the posterior itself is formally incorrect if non-Gaussianities exist,
since it incorporates a prior that assumes otherwise; the correct way to include
them would be to forward model them in some form as part of the prior.

Equation \eqref{eq:lnP} is the posterior probability in terms of the unlensed field.
The probability can also be parametrized in terms of the lensed field, $\tilde
f$, which introduces an additional Jacobian term $\partial f / \partial \tilde f
=\Lphi{\phi}^{-1}$ from the change of variables,
\begin{align}
-2&\log \, \mathcal P(\tilde f,\phi,\theta \,|\, d)=\nonumber\\
&= -2\log \mathcal  P(\Lphi{\phi}^{-1} \tilde f,\phi,\theta \,|\, d) + 2\log\left|\det \Lphi{\phi}\right| \nonumber\\
&= (d-\tilde f)^\dagger \Cn^{-1} (d-\tilde f) \label{eq:lnPlensed}\\
&\qquad +\, \tilde f^\dagger \Lphi{\phi}^{-\dagger} \,\Ccmbtheta^{-1}\, \Lphi{\phi}^{-1} \tilde f\,  + \,  \phi^\dagger \,\Clentheta^{-1}\,\phi  \nonumber  \\
&\qquad +  \logdet\Clentheta + \logdet\Ccmbtheta  + 2 \log\left|\det\Lphi{\phi}\right|  \nonumber
\end{align}

The difficulty is that one now needs to know the absolute value of the lensing
determinant, $\left| \det \Lphi{\phi} \right|$, which cannot otherwise be
ignored since it depends on one of the arguments of the probability function. In
Sec.~\ref{sec:lensdet} we will show that in the limit of infinite resolution,
this determinant is equal to unity, but on pixelized maps can differ from one
depending on the pixelized lensing approximation in use. Indeed, for the
standard Taylor series expansion for lensing, we will show that the determinant
cannot be treated as constant with respect to $\phi$. Our solution is to develop
a new pixelized lensing approximation, which we call \textsc{LenseFlow}, and
which always has determinant equal to unity even on pixelized maps. For now we
will continue the discussion, and delay a description of \textsc{LenseFlow}
until Sec.~\ref{sec: lenseflow}.

\subsection{Relation to marginalized posteriors}
\label{sec:margnalized_posterior}

In studies where the parameter of interest is $\phi$, one may integrate out the
unknown $f$ to obtain the marginal posterior given by 
\begin{align}
	\label{eq:marginal_post}
    \mathcal P(\phi \,|\, d) = \int \!\mathrm d f \, \mathcal P(f, \phi \,|\,d).
\end{align}
(we will drop explicitly labeling $\theta$ in this section).

This integral can be done analytically, and it is this probability distribution
which is maximized by the algorithms given in \cite{hirata2003} and
\cite{carron2017}. In this section we compare the differences between this
marginal estimate and the one developed here which maximizes the joint $\mathcal
P(f,\phi\,|\, d)$.

The analytic marginalization over $f$ can be regarded as an application of the
Laplacian approximation method, which is exact in this case due to the
Gaussianity of $\mathcal P(f\,|\,\phi, d)$, and which we give here since it also
helps clarify the differences between the two estimates and the algorithms for
computing them. To derive the Laplace approximation, first notice that for any
fixed $\phi$ the function $f\,{\mapsto}\,\log\mathcal P(f,\phi\,|\,d)$ is
quadratic in $f$. This implies there exists a normalization, $Z(\phi)$,
which makes $f\,{\mapsto}\,\mathcal P(f,\phi\,|\,d)/Z(\phi)$ a Gaussian
probability measure. In particular there exists $\hat f(\phi)$ and
$\Sigma({\phi})$ such that, up to a constant, 
\begin{align}
-2&\log\big[\mathcal P(f,\phi\,|\,d)/Z(\phi)\big] = \nonumber\\
&= [f-\hat f(\phi)\big]^\dagger \Sigma(\phi)^{-1}\big[f-\hat f(\phi)\big] + \log\det \Sigma(\phi)
\end{align}
where $\hat f(\phi) = \argmax_f \log\mathcal P(f,\phi\,|\,d)$ and $\Sigma(\phi)$
is the negative inverse Hessian of $f\mapsto \log\mathcal P(f,\phi\,|\,d)$. One
can explicitly compute $\Sigma(\phi)$, $\hat f(\phi)$ and $Z(\phi)$ as follows
\begin{align}
	\Sigma(\phi) &= \big[\Lphi{\phi}^\dagger\,  \Cn^{-1} \Lphi{\phi} + \Ccmb^{-1}   \big]^{-1} \\
	\hat f(\phi) &= \Sigma(\phi) \,\Lphi{\phi}^{\dagger}\, \Cn^{-1} d \\
	Z(\phi)  &= \det\Sigma(\phi)^{\frac{1}{2}}\,\mathcal P(\hat f(\phi),\phi\,|\,d)
\end{align}
By multiplying and dividing $Z(\phi)$ in \eqref{eq:marginal_post}, while using
the fact that $\mathcal P(f,\phi\,|\,d)/Z(\phi)$ integrates to $1$ over $f$, the
marginal posterior over $\phi$ is then given by
\begin{align}
\mathcal P(\phi \,|\, d)
&=  \det\Sigma(\phi)^{\frac{1}{2}}\,\mathcal P(\hat f(\phi), \phi\,|\,d) \nonumber\\
&\propto \frac{\mathcal P(\hat f(\phi),\phi \,|\,d)}{\det\big[\Lphi{\phi}\,  \Ccmb \Lphi{\phi}^\dagger + \Cn\big]^{\frac{1}{2}}} \label{eq:laplace_approx}
\end{align}
Equation \eqref{eq:laplace_approx} thus shows the marginal posterior on $\phi$
in the form of the Laplace approximation. 

Now, to distinguish marginal versus joint MAP estimates we set the following
notation
\begin{align}
	\hat \phi_{M}  &\equiv \argmax_\phi \, \mathcal P(\phi \,|\, d) \label{eq: phi_M def} \\%\, \frac{\mathcal P(\hat f(\phi), \phi \,|\,d)}{\det\big[\Lphi{\phi}\,  \Ccmb \Lphi{\phi}^\dagger + \Cn \big]^{\frac{1}{2}}} \\
	\hat \phi_{J}  &\equiv \argmax_\phi \, \mathcal P(\hat f(\phi), \phi \,|\, d) \label{eq: phi_J def}\\
	\hat f_{M} &\equiv \hat f({\hat \phi_{M}})\quad\text{ and }\quad\hat f_{J}\equiv \hat f({\hat \phi_{J}})
\end{align}
where $\hat \phi_{M}$ corresponds to the marginal estimate of $\phi$ and
\begin{align*} (\hat \phi_J, \hat f_J) &= \argmax_{\phi, f} \, \mathcal
P(f,\phi\, | \, d) \end{align*}
corresponds to the joint MAP estimate of both $\phi$ and $f$.

First notice that $\hat \phi_{M}$ and $\hat \phi_{J}$ are maximizing
non-trivially different objectives, \eqref{eq: phi_M def} versus \eqref{eq:
phi_J def}, so clearly $\hat \phi_{J}\neq \hat \phi_{M}$ and hence $\hat
f_{J}\neq \hat f_{M}$ as well. The fact that these estimates are different is an
explicit manifestation of the non-Gaussianity of the posterior $\mathcal
P(f,\phi\,|\, d)$, for otherwise marginal and joint MAP estimates would agree.
More importantly, however, $\hat f_{M}$ can not be interpreted as a MAP estimate
of the CMB, but rather as an intermediate variable used for the Laplace
approximation technique of marginalization. This is not to say that $\hat f_{M}$
could not have reasonable sampling properties as a statistical estimator, but
rather that $\hat f_{M}$ does not have an interpretation in the Bayesian
framework.

In Section \ref{sec:maximization} we present an iterative algorithm for
computing $(\hat \phi_{J}, \hat f_{J})$ which shares some similarities to the
one given in \cite{carron2017} for computing $\hat \phi_{M}$. However, the
similarities are largely superficial. While both algorithms do generate a
sequence of iterations $\ldots, (f_i, \phi_i), \ldots$ where $f_{i}$ is defined
recursively by a generalized Wiener filter of the unlensed CMB given the
previous $\phi_{i-1}$, i.e.\! $f_{i} = \hat f({\phi_{i-1}})$, important
differences arise in how $\phi_{i}$ is computed. In \cite{carron2017} the update
$\phi_{i}$ is computed as the solution to a stationary equation characterizing
the maximum of \eqref{eq:laplace_approx} with $f_i$ in place of $\hat f(\phi)$.
In contrast, the algorithm given in Section \ref{sec:maximization} updates
$\phi_i$ using the lensed CMB parameterization $(\phi,\tilde f\,)$ and, as such,
is computed as an approximate maximizer of the {\em lensed} posterior given
$\tilde f_i = \mathcal L({\phi_{i-1}}) f_i = \mathcal L({\phi_{i-1}})\hat
f({\phi_{i-1}})$. In particular, 
\begin{align}
    \phi_{i} \approx \argmax_\phi \mathcal P(\Lphi{\phi}^{-1} \tilde f_i, \phi\,|\, d). 
\end{align}
One way to see the impact of this difference is through the
data term $-\frac{1}{2}(d-\tilde f_i)^\dagger \Cn^{-1} (d-\tilde f_i)$,
appearing  in $\log\mathcal P(\Lphi{\phi}^{-1} \tilde f_i, \phi \,|\, d ) $,
which is completely invariant to changes in $\phi$. This allows our algorithm to
make large jumps in $\phi$ that are completely de-coupled from the data and
experimental conditions. Notice that this property also extends to posterior
sampling and results in fast mixing Gibbs iterations. Indeed, this subtle
difference gives a succinct way to see the key advantage gained when working
with the lensed parameterization $(\phi, \tilde f)$ versus unlensed
parameterization $(\phi, f)$.

All of this raises the question: which estimate should one use,  $\hat \phi_M$
or $\hat \phi_J$? Technically, neither $\hat \phi_M$ nor $\hat \phi_J$ is \lq\lq
optimal\rq\rq, at least with respect to posterior expected quadratic error (the
marginal expected value being optimal). We will see in Section
\ref{sec:results_masking} that there are some apparent advantages to working
with $\hat \phi_M$ in that the extra determinant term in
\eqref{eq:laplace_approx} automatically removes a \lq\lq mean field\rq\rq  which
becomes large in the presence of pixel space masking.  However, the real goal of
a Bayesian analysis is quantification of uncertainty and in that respect, MAP
estimates are usually of limited scientific use. When considering the full
problem of posterior sampling, the extra determinant term in $\hat \phi_M$ now
becomes a difficult computational obstacle for sampling algorithms. Moreover,
the joint $\mathcal P(f, \phi\,|\, d)$ has the advantage of simultaneously
characterizing both the delensed CMB marginal $\mathcal P(f\,|\, d)$ as well as
$\mathcal P(\phi\,|\, d)$.

%%%%%%%%%%%% section %%%%%%%%%%%%%%%%
\section{The maximization algorithm}
\label{sec:maximization}

%%%%%%%%%%%%%%%%%%%%% BEGIN figure %%%%%%%%%%%%%%%%%%%%%
\begin{figure*}
\begin{center}
\includegraphics[width=\textwidth]{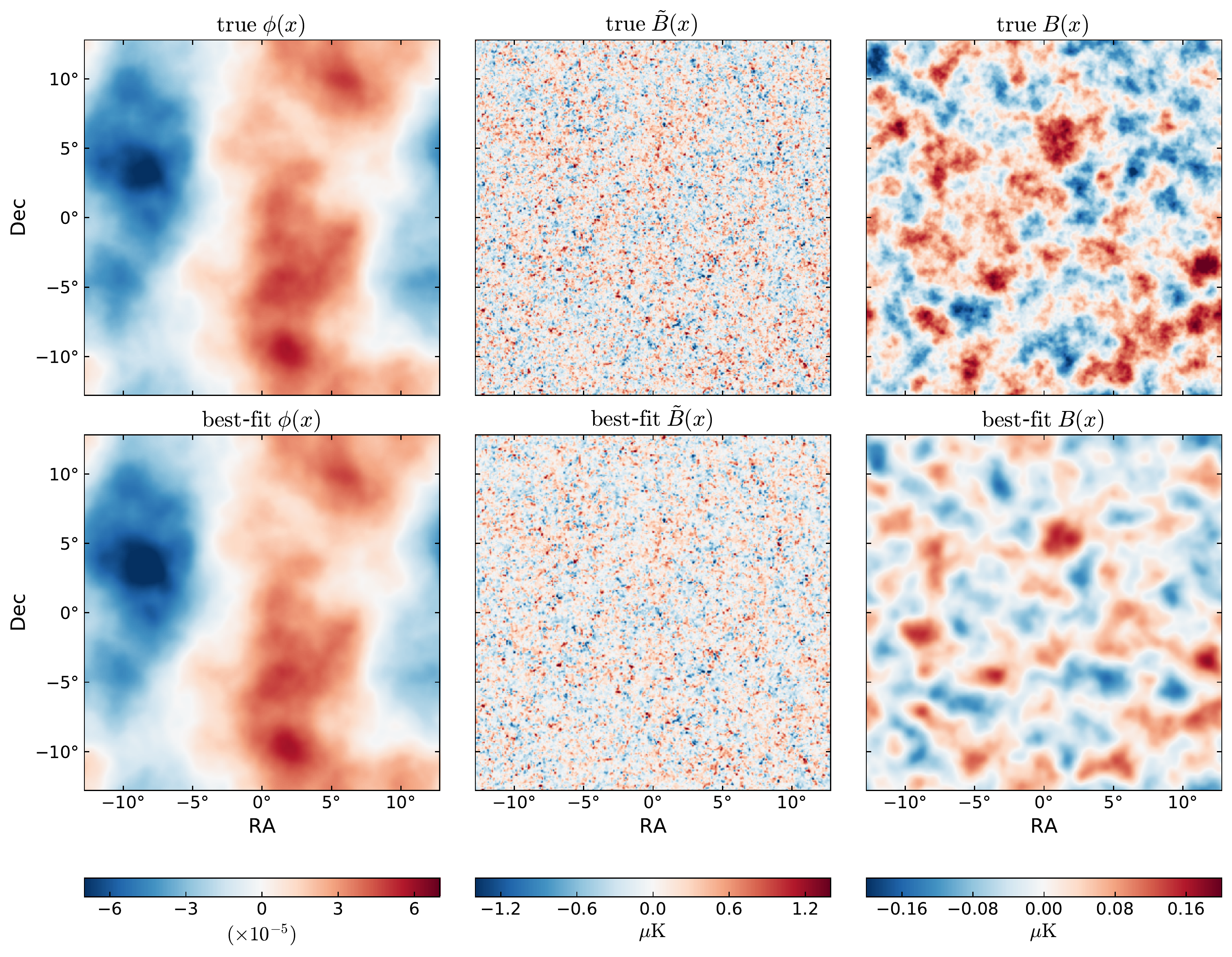}
\end{center}

\caption{The reconstructed $\phi$ and lensed/unlensed $B$ maps from a run of our
algorithm on simulated data (bottom row), as compared to the simulation truth
(top row). This is for the run with only Fourier-space masking described in
Sec.~\ref{sec:results_nomasking}. The reconstruction, as expected, resembles a
Wiener filter solution wherein low signal-to-noise modes are attenuated. }

\label{fig:nomasking_maps}
\end{figure*}
%%%%%%%%%%%%%%%%%%%%% END figure %%%%%%%%%%%%%%%%%%%%%

With the target probability function \eqref{eq:lnPlensed} in hand, we now
describe our maximization algorithm. We have attempted a number of different
approaches, but the most efficient we have found is based on the observation
that maximizing separately with respect to $\tilde f$ and to $\phi$ cleanly
breaks the problem up into two simple pieces, a Wiener filter and something
which is independent of the instrument and data. To that end, we employ a
coordinate descent, i.e. alternating maximization steps in the $\tilde f$ and
$\phi$ directions separately. Coordinate descent also has the advantage that it
is essentially the maximization analog to Gibbs sampling, which is exactly the
sampling algorithm shown successful for temperature in \cite{anderes2015}. We
therefore expect the developments that we present here which make the
maximization workable for polarization to also transfer to the sampling case.

Consider first the coordinate descent step for $\tilde f$. The maximum
probability for $\tilde f$ given fixed $\phi$ can be calculated by taking the
gradient of the likelihood,
\begin{align}
\frac{\partial}{\partial \tilde f} \log\mathcal P(\tilde f, \phi\,|\,d) &=(d -\! \tilde f)^{\dagger}  \Cn^{-1} - \tilde f^{\,\dagger}\Lphi{\phi}^{-\dagger} \Ccmb^{-1} \Lphi{\phi}^{-1}
\end{align}
and setting it to zero. This gives an explicit solution,
\begin{align}
\tilde f  &= \Lphi{\phi} \left[\Ccmb^{-1} +  \Lphi{\phi}^{\dagger}\,  \Cn^{-1} \Lphi{\phi} \right]^{-1} \!\Lphi{\phi}^{\dagger}\, \Cn^{-1} d \label{ea: iterative eq f precond}
\end{align}
which can be recognized as an ordinary Wiener filtering of the data with a
$\phi$-dependent signal covariance. The challenge is inverting the quantity in
brackets in \eqref{ea: iterative eq f precond}. We find that inverting it with a
simple preconditioned conjugate gradient (with a preconditioning matrix that
assumes $\phi=0$ and noise which is diagonal in Fourier space) works
sufficiently well. The reduction of part of the problem to the well known Wiener
filter problem is a major advantage of the coordinate descent, since many Wiener
filter algorithms exist which are efficient and can be guaranteed to converge,
unlike generic non-linear optimization algorithms.

Now consider the coordinate descent step for $\phi$. Here the gradient is given
by,
\begin{align}
 \frac{\partial}{\partial \phi} &\log\mathcal P(\tilde f, \phi\,|\,d) \nonumber\\
 &= -\frac{1}{2} \frac{\partial}{\partial \phi} \left[ \tilde f^{\,\dagger}\Lphi{\phi}^{-\dagger} \Ccmb^{-1} \Lphi{\phi}^{-1}\tilde f \right] - \phi^{\dagger} \Clen^{-1} \nonumber\\
 % &=  -\frac{1}{2} \frac{\partial \left[ \tilde f^{\,\dagger}\Lphi{\phi}^{-\dagger} \Ccmb^{-1} \Lphi{\phi}^{-1}\tilde f \right] }{\partial (\Lphi{\phi}^{-1}\tilde f)}\left[ \tfrac{\partial }{\partial \phi} \Lphi{\phi}^{-1}\tilde f \right]  - \phi^{\dagger} \Clen^{-1}
% \nonumber\\
&= -\tilde f^{\,\dagger}\Lphi{\phi}^{-\dagger} \Ccmb^{-1} \left[ \tfrac{\partial }{\partial \phi} \Lphi{\phi}^{-1}\tilde f \right] - \phi^{\dagger} \Clen^{-1}.
\end{align}
Taking the adjoint and setting to zero yields,
\begin{align}
\left[ \tfrac{\partial }{\partial \phi} \Lphi{\phi}^{-1}\tilde f \right]^\dagger \Ccmb^{-1} \Lphi{\phi}^{-1} \tilde f - \Clen^{-1} \phi = 0,
\label{ea: iterative eq phi}
\end{align}
Unlike the $\tilde f$ step, it is not possible to obtain an explicit solution
for $\phi$. Instead, we solve this iteratively with a quasi Newton-Raphson step,
\begin{align}
\phi_{i+1} = \phi_i - \alpha \mathcal{H}(\tilde f, \phi_i)^{-1} \,\frac{\partial}{\partial \phi} &\log\mathcal P(\tilde f, \phi_i\,|\,d)
\label{ea: iterative eq phi nr}
\end{align}
Here $\mathcal{H}(\tilde f, \phi_i)$ denotes the Hessian of $\phi \mapsto
\mathcal{P}(\tilde f, \phi\,|\, d)$  and $\alpha$ is a scalar coefficient over
which we perform a line-search to maximize the probability. We take
$\mathcal{H}\approx\Clen$, which is the contribution to the Hessian from only
the $\phi$-prior term, but which we find works extremely well in practice. By
the time we are close to maximum, we expect a single Newton-Raphson step would
take us quite close to the exact solution of \eqref{ea: iterative eq phi}, but
we have found that even before we reach the maximum we can get away with just a
single iteration of \eqref{ea: iterative eq phi nr} at each coordinate descent
step and convergence is still quite fast.

For the $\phi$ step, the coordinate descent has removed all explicit dependence
on the instrument; note that neither the data nor the noise covariance (and
hence no masking, transfer function, etc...) appear explicitly in \eqref{ea:
iterative eq phi}. It is worth re-stating that this would {\it not} have been
the case if we were performing coordinate descent with respect to $(f,\phi)$ as
opposed to  $(\tilde f,\phi)$, hence this can be seen as another fundamental
advantage of the lensed parametrization. 

The maximization algorithm then simply starts at $\phi\,{=}\,0$ and alternates
these two coordinate descent steps, until acceptable convergence is reached.
There is only one additional detail we need to describe which is necessary for
convergence to happen efficiently enough, and that is our use of a cooling
schedule for the covariance, $\Ccmb$. By this we mean that we replace $\Ccmb$
everywhere that it appears in the iterating equations with a new covariance,
which we call the cooling covariance and denote with $\preCcmb{\hat}$. It is
initially set to the {\it lensed} CMB covariance (which we will denote by
$\preCcmb{\tilde}$), then progressively ``cooled'' it towards $\Ccmb$. By the
final iteration we cool to exactly $\Ccmb$ and thus are maximizing the true
posterior.

The cooling scheme is aimed at keeping the power-spectrum of the $\tilde f$
estimate constant across iterations and roughly matching the expected
power-spectrum of the lensed CMB. This happens at the expense of making the
power-spectrum of $f$ not always match the unlensed spectrum, but is
advantageous nevertheless since it is in the lensed parameterization that we are
performing the coordinate descent. To achieve this goal, the cooling scheme
takes $\preCcmb{\hat}$ at a given iteration to be the expected power-spectrum of
the true lensed field delensed by the current $\phi$ estimate at that iteration.
For a given configuration (i.e. noise level, pixelization, map size, etc...), we
can calculate this covariance with simulations, since we have access to the true
lensed field. In fact, we find that only one simulation is necessary, as we can
greatly reduce sample variance fluctuations by modeling the cooling covariance
as a geometric mean between the lensed and unlensed $C_\ell$'s, with an
$\ell$-dependent weight, $w_\ell$, and heavily interpolating this quantity based
on the observed $BB$ spectrum of the one simulation. This produces a set of
geometric weights $w_\ell^i$ for each iteration $i$ which we use in subsequent
runs. These weights, along with the data and the number of iterations are the
only inputs to the maximization procedure, which we summarize in Algorithm
\ref{maxalg} below.

% EA: I needed the addition of `[H]` to get latex to compile.
% Ref: https://tex.stackexchange.com/questions/231191/algorithm-in-revtex4-1
\begin{algorithm}[H]
\caption{Joint posterior maximization}\label{maxalg}
\begin{algorithmic}[1]
\Procedure{JointPosteriorMax}{$d,N,w_\ell^i$}
\State $\phi_1=0$, $f_1=0$, $\tilde f_1=0$ $\vphantom{\big[^\dagger}$
\For{$i=1...N-1$ $\vphantom{\big[^\dagger}$ }
\State $\preCcmbpost{\hat}{\ell} \,=\,(\preCcmbpost{}{\ell})^{w_\ell^i}  (\preCcmbpost{\tilde}{\ell})^{1-w_\ell^i}$ $\vphantom{\big[^\dagger}$ 
\State $A = \preCcmb{\tilde}^{-1} +  \Lphi{\phi_i}^{\dagger}  \Cn^{-1} \Lphi{\phi_i}$ $\vphantom{\big[^\dagger}$
\State $b = \Lphi{\phi_i}^{\dagger} \Cn^{-1} d$ $\vphantom{\big[^\dagger}$
\State $f_{i+1} = A^{-1}b$ $\vphantom{\big[^\dagger}$ \Comment{Solve via CG}
\State $\tilde f_{i+1} = \Lphi{\phi_i} f_{i+1}$ $\vphantom{\big[^\dagger}$
\State $g = \left[ \tfrac{\partial}{\partial \phi} \Lphi{\phi_i}^{-1} \tilde f_{i+1} \right]^\dagger \!\preCcmb{\tilde}^{-1}\, f_{i+1} + \Clen^{-1} \phi_i$
\State $\alpha = \text{\sf Max}_{\alpha}\,\mathcal{P}\big(\tilde f_i,\, \phi_i - \alpha\, \Clen g\,|\,d\big)$  $\vphantom{\big[^\dagger}$

\State $\phi_{i+1} = \phi_i - \alpha \, \Clen g$ $\vphantom{\big[^\dagger}$
\EndFor $\vphantom{\big[^\dagger}$
\State \textbf{return} $\phi_N, f_N, \tilde f_N$ $\vphantom{\big[^\dagger}$
\EndProcedure $\vphantom{\big[^\dagger}$
\end{algorithmic}
\end{algorithm}

We have already ascertained in the previous section that the lensing operation
which appears throughout the algorithm, or more specifically its inverse, needs
to be area-preserving. Thus a requirement on the lensing algorithm which we use
is, 
\begin{enumerate}[label=\textbf{\arabic*})]
    \item\label{ea: LF det} $\left|\det\big(\Lphi{\phi}^{-1}\big)\right| = 1$ to numerical precision;
\end{enumerate}
Examining Algorithm \ref{maxalg}, we note that we also need two other things of
the lensing operation, 
\begin{enumerate}[label=\textbf{\arabic*})]
    \setcounter{enumi}{1}
    \item\label{ea: LF transpose} Computation of $\Lphi{\phi}^\dagger f$
    \item\label{ea: LF gradient} Computation of $\big[\frac{\partial}{\partial \phi} \Lphi{\phi}^{-1} \tilde f\, \big]^\dagger$
\end{enumerate}
In the next section we develop \textsc{LenseFlow} which performs pixelized
lensing in a way that simultaneously satisfies \ref{ea: LF det}, \ref{ea: LF
transpose},  and \ref{ea: LF gradient} above.

%%%%%%%%%%%% section %%%%%%%%%%%%%%%%
\section{Lense Flow} \label{sec: lenseflow}

\textsc{LenseFlow} is an algorithm that utilizes an ordinary different equation
(ODE) to describe the lensing operator, $\Lphi{\phi}$.  An auxiliary ``time''
variable is introduced which continuously connects the lensed and unlensed maps
such that $\Lphi{\phi} f$ is given by the solution of an ODE over map pixels
with initial conditions $f$. Because the ODE is homogenous, we can regard the
pixel values as ``flowing'' from their unlensed values to their lensed ones,
hence the name \textsc{LenseFlow}. There are a number of advantages one obtains
with an ODE characterization of a linear operator. First, operator inversion
simply corresponds to running the ODE in reverse. Secondly, log determinants can
be analyzed using the trace of the velocity operator, integrated over time.
Finally, in many cases higher order derivatives with respect to both initial
conditions and parameters of the ODE have their own ODE characterizations. In
the case of \textsc{LenseFlow}, these enable fast and accurate calculation of
gradient and Hessian operators of  $\log \mathcal P (\tilde f, \phi\,|\, d)$
with respect to both $\tilde f$ and $\phi$.\footnote{Incidentally, the ODEs for
calculating these derivatives are exactly analogous to the backpropagation
techniques used for learning deep neural networks \citep{caterini2016} but are
derived here completely from ODE theory.}

We begin to define \textsc{LenseFlow} by introducing an artificial time variable
to the CMB field which connects the lensed CMB at $t\,{=}\,1$ with the unlensed
CMB at $t\,{=}\,0$. In particular, for $t\in [0,1]$ let
\begin{align}\label{ea: def of f_t}
    f_t(x) \equiv f(x+t\nabla\phi(x))
\end{align}
so that $f_0(x) = f(x)$ and  $f_1(x) = \tilde f(x)$. An ordinary differential
equation for $f_t$ can be derived from
\begin{align}
    \frac{d f_t(x) }{dt} = \nabla^i f(x + t\nabla \phi(x)) \, [\nabla \phi(x)]^i
\end{align}
and the following chain rule
\begin{align}
	\nabla^i f_t(x) =   \nabla^j f(x + t\nabla \phi(x)) \, \big[\delta^{ij} + t \nabla^{i} \nabla^{j} \phi(x) \big]
    \label{eq:chainrule}
\end{align}
where $\nabla^i\equiv\partial/\partial x^i$ (we are working here in the flat-sky
approximation) and $\delta^{ij}$ is the Kronecker delta. The quantity in
brackets in \eqref{eq:chainrule} represents the $2\times 2$ Jacobian of the map
$x\mapsto x + t\nabla \phi(x)$, which for $t\,{=}\,1$ is often called the
magnification matrix; we will henceforth label it with $M_t$. It is invertible
in the weak lensing regime in which we work here, thus we can combine the above
two equations to yield that $f_t$ satisfies
\begin{align}\label{eq:lenseflow_ode}
    \dot f_t = (\nabla^j \phi)\, (M_t^{-1})^{ji}\, \nabla^i f_t.
\end{align}
By definition, solving the ODE \eqref{eq:lenseflow_ode} forward in time, $t=0
\rightarrow 1$, represents the lensing operation. Moreover, exact inverse
lensing simply corresponds to flowing the ODE backwards in time, $t=1
\rightarrow 0$. Notice that invertibility of \textsc{LenseFlow} also extends to
discrete pixel-to-pixel lensing by replacing the gradient, $\nabla$, in
\eqref{eq:lenseflow_ode},  with its discrete Fourier analog.

The fact that  \textsc{LenseFlow} is an area preserving linear operator, i.e.\!
that \ref{ea: LF det} holds, follows directly from \eqref{eq:lenseflow_ode}.  To
see why, first define
\begin{align}
	p_t^i & = (\nabla^j \phi) (M_t^{-1})^{ji}\,   \label{ea: p def}
\end{align}
so that \eqref{eq:lenseflow_ode} is written in compact form $\dot f_t = p_t^i\,
\nabla^i f_t$. Now since the flow from $f_0$ to $f_1$ can be written as
composition of infinitesimally small linear operations, the lensing operator
$\Lphi{\phi}$ is decomposed as follows
\begin{align}\label{eq:lenseflow_discrete}
    f_1 &= \underbrace{\left[1+\epsilon\,\, p_{t_n}^i\! \nabla^i\, \right]\cdots \left[1+\epsilon \,\, p_{t_0}^i\! \nabla^i\, \right]}_{=\Lphi{\phi}} f_0
\end{align}
where %$p_t =  M_t^{-1} \cdot \nabla \phi$,
$\epsilon = \frac{1}{n} =
t_{i+1} - t_i$ and $t_0 = 0$. Notice that
\begin{align}\label{eq:lenseflow_det}
    \log\det\left[1+\epsilon\,\, p_{t}^i\, \nabla^i\,  \right]
    &= \epsilon\, \text{Tr}\left[\, p_{t}^i\, \nabla^i\, \right] + \mathcal O(\epsilon^2) \nonumber \\
&= \mathcal O(\epsilon^2)
\end{align}
where the last equality follows since the operator $\nabla^i$ is Hermitian
anti-symmetric. This applies also to the inverse operation, thus up to ODE
time-step discretization error, condition \ref{ea: LF det} holds for
\textsc{LenseFlow}, independent of pixel size, 
\begin{align}\label{eq:lenseflow_det}
    \lim_{\epsilon\to0} \det\big(\Lphi{\phi}^{-1}\big) = 1
\end{align}

It will be useful to have a compact notation for the decomposition of a linear
operator characterized by an ODE, as in \eqref{eq:lenseflow_discrete}. To that
end define
\begin{align}
    \underset{t = t_0\rightarrow t_n}{\sf ODE}\!\!\big\{ V_t \big\} \equiv \left[1+\epsilon\, V_{t_n}\, \right]\cdots \left[1+\varepsilon \, V_{t_0}\, \right]
\end{align}
where $V_t$ represents a ``velocity operator'' generating an ODE of the form
$\dot f_t = V_t f_t$ and where $\varepsilon = t_{i+1}-t_{i}$ represents an
infinitesimal time step for an ordered equidistant sequence of time points $t_0,
t_1, \ldots, t_n$. This allows us to succinctly define \textsc{LenseFlow} as,
\begin{align}
    %\boxed{
	\Lphi{\phi} = \underset{t=0\rightarrow1}{\sf ODE}\big\{ p_t^i\, \nabla^i \big\}.
	%}
    \label{eq:lenseflow}
\end{align}
The infinitesimal ODE expansion also makes it clear that both the inverse and
adjoint of an ODE operator is also an ODE operator, but with time reversed, and
in the latter case with a negative adjoint velocity
\begin{align}
	&\Big[\underset{\,t=t_0\rightarrow t_n}{\sf ODE}\!\!\left\{ V_t \right\}\Big]^{-1} = \underset{t=t_n\rightarrow t_0}{\sf ODE}\!\big\{V_t \big\}
	\label{eq:ode_inverse}\\
    &\Big[\underset{\,t=t_0\rightarrow t_n}{\sf ODE}\!\!\left\{ V_t \right\}\Big]^\dagger \,\,\,= \underset{t=t_n\rightarrow t_0}{\sf ODE}\!\big\{\!-V_t^\dagger \big\}.
    \label{eq:ode_adjoint}
\end{align}
Due to the fact that $[\, p_t\cdot \nabla\,]^\dagger \!f = -\nabla^i (p_{t}^i
f)$, the latter equation can be used to compute the adjoint lensing operator
\begin{align}
    %\boxed{
	\Lphi{\phi}^\dagger = \underset{t=1\rightarrow0}{\sf ODE}\left\{ \nabla^i ( p^i_{t}\,\text{\small$\bullet$}) \right\}
	%}
    \label{eq:adjointflow}
\end{align}
where the expression $\nabla^i ( p^i_{t}\,\text{\small$\bullet$})$ is shorthand
for the operator $f\mapsto \nabla^i (p_{t}^i\, f)$. Notice that
\eqref{eq:adjointflow} achieves \ref{ea: LF transpose}, another of our
requirements for the lensing operation. Although not explicitly needed, note
also that the operator $\Lphi{\phi}^{-\dagger}$ is conveniently computed by
simply applying a time reversal of \eqref{eq:adjointflow}, as per
\eqref{eq:ode_inverse}.

For the final requirement in \ref{ea: LF gradient}, we need to compute
derivatives of the inverse lensing operator with respect $\phi$ and initial
condition, $f_0$. Introducing infinitesimal perturbations $\delta \phi$ and
$\delta\! f_t$ into \eqref{eq:lenseflow_ode}, we have
\begin{align}
	\dot {\delta\! f_t}  &=
 	(\nabla^i \delta\phi)\,  (M_t^{-1})^{ij} \nabla^j f_t + (\nabla^i \phi) \, \delta (M_t^{-1})^{ij} \nabla^j f_t  \nonumber \\
	&\qquad + (\nabla^i \phi)\,  (M_t^{-1})^{ij} \nabla^j \delta\! f_t
\label{eq:deltaflow_ode}
\end{align}
Simplifying $\delta( M_t^{-1} )^{ij}$ and treating $\delta\phi$ as a time dependent
variable results in
\begin{equation}\label{ea: delta flow A}
\Bigg[\!\!
\begin{array}{c}
    \dot {\delta\! f_t}\\
    \dot {\delta \phi_t}
\end{array}
\!\Bigg]
=
\Bigg[\!\!
\begin{array}{cc}
    p_t^i\, \nabla^i\,\,  &  \,\, v_t^i\, \nabla^i -t W_t^{i j} \nabla^i \nabla^j \\
    0 & 0
\end{array}
\!\!\Bigg]
\Bigg[\!\!
\begin{array}{c}
    {\delta\! f_t}\\
    {\delta \phi_t}
\end{array}
\!\Bigg]
\end{equation}
where $p_t$, $v_t$, and $W_t$ are defined by
\begin{align}
    p_t^i & = (\nabla^j \phi) (M_t^{-1})^{ji}\,   \label{ea: p def} \\
    v_t^i & = (\nabla^j f_t)\, (M_t^{-1})^{ji}  \label{ea: delta flow B}\\
    W_t^{i j} & = (\nabla^p \phi) \, (\nabla^q f_t) \, (M_t^{-1})^{p i}\, (M_t^{-1})^{j q}.\label{ea: delta flow C}
\end{align}
(the definition of $p_t$ is repeated here for clarity). It is important to note
that, unlike $p_t^i$ which is a scalar field for each index $i$, the quantities
$W_t^{i j}$ and $v_t^i$ are instead a TQU vector of temperature and polarization
fields at each index. As is usually implicitly assumed, multiplication between a
scalar field and a TQU vector broadcasts over the TQU indices. One important
consequence of this is that the adjoint of $ W_t^{i j} \nabla^i \nabla^j$ and
$v_t^i\, \nabla^i$ are given by $\nabla^j \nabla^i ((W_t^{i j})^\mathsf{T}
\,\text{\small$\bullet$} )$ and $-\nabla^i ((v_t^i)^\mathsf{T}
\,\text{\small$\bullet$}) $, respectively, where we define $\mathsf{T}$ to
represent a transpose of just the TQU indices. For example, if $f$ is a TQU
vector of fields, $f^\mathsf{T} f$ represents the scalar field $I^2 + Q^2 + U^2$
(in contrast to $f^\dagger f$, for example, which would be a single number).

If we now consider a map between the lensed and unlensed parametrizations,
$(f,\phi)\mapsto(\tilde f,\phi)$, the Jacobian $\mathbf{J}\equiv \frac{\partial
(\tilde f,\phi)}{\partial (f,\phi)}$ and its inverse are given by
\begin{align}
    \mathbf{J}
	= \left[\begin{array}{cc}
        \dfrac{\partial \tilde f}{\partial f}\, & \,\dfrac{\partial \tilde f}{\partial \phi} \\
        0 & 1
    \end{array}\right]\;\;\;\;\;
    \mathbf{J}^{-1}
	= \left[\begin{array}{cc}
        \dfrac{\partial f}{\partial \tilde f}\, & \,\dfrac{\partial f}{\partial \phi} \\
        0 & 1
    \end{array}\right]
\end{align}
Equations \eqref{ea: delta flow A}-\eqref{ea: delta flow C} show that
$\mathbf{J}$ can be computed as,
\begin{align}
    %\boxed{
	\mathbf{J} = \underset{t=0\rightarrow1}{\sf ODE}
    %\resizebox{.65\hsize}{!}{$
	\left\{\!\!\Bigg[\!\!\begin{array}{cc}
        p_t^i \, \nabla^i\,\,  &  \,\,v_t^i\, \nabla^i - t W_t^{i j} \nabla^i \nabla^j \\
        0 & 0
    \end{array}\!\!\Bigg]\!\!\right\}
	%$}
	%}
    \label{eq:deltaflow}
\end{align}
and \eqref{eq:ode_adjoint} immediately gives that the adjoint Jacobian is
\begin{align}
    %\boxed{
	\!\mathbf{J}^\dagger = \underset{t=1\rightarrow0}{\sf ODE}
    %\resizebox{.65\hsize}{!}{$
	\left\{\!\!\Bigg[\!\!\begin{array}{cc}
        \nabla^i (p_t^i\,\text{\small$\bullet$}) & 0 \\
        \nabla^i ((v_t^i)^\mathsf{T}\,\text{\small$\bullet$}) \!+\! t \nabla^j \nabla^i ((W_t^{i j})^\mathsf{T}\,\text{\small$\bullet$})\,  & \,0
    \end{array}\!\!\Bigg]\!\!\right\}
	%$}
	%}
    \label{eq:deltatransflow}
\end{align}
Note that the velocities for the Jacobian ODE depend on $f_t$, which can be
precomputed from an initial application of the corresponding lensing operator,
or in some cases simply solved for in unison.

As before, the inverse of \eqref{eq:deltatransflow} can be trivially computed by
time reversal of the ODE, using \eqref{eq:ode_inverse}. The bottom left block of
$\mathbf{J}^{-\dagger}$ then satisfies
\[
\mathbf{J}^{-\dagger}
\Bigg[\!\!
\begin{array}{c}
    {\delta\! f}\\
    { 0}
\end{array}
\!\Bigg] =
\left[\!\!
\begin{array}{c}
    * \\
    \big[\tfrac{\partial}{\partial \phi} \Lphi{\phi}^{-1} \tilde f\, \big]^\dagger \delta\! f
\end{array}
\!\right]
\]
which is exactly the necessary derivative which satisfies the final requirement
of \ref{ea: LF gradient}.

%%%%%%%%%%%%%%%%%%%%% BEGIN figure %%%%%%%%%%%%%%%%%%%%%
\begin{figure*}
\begin{center}
\includegraphics[width=\textwidth]{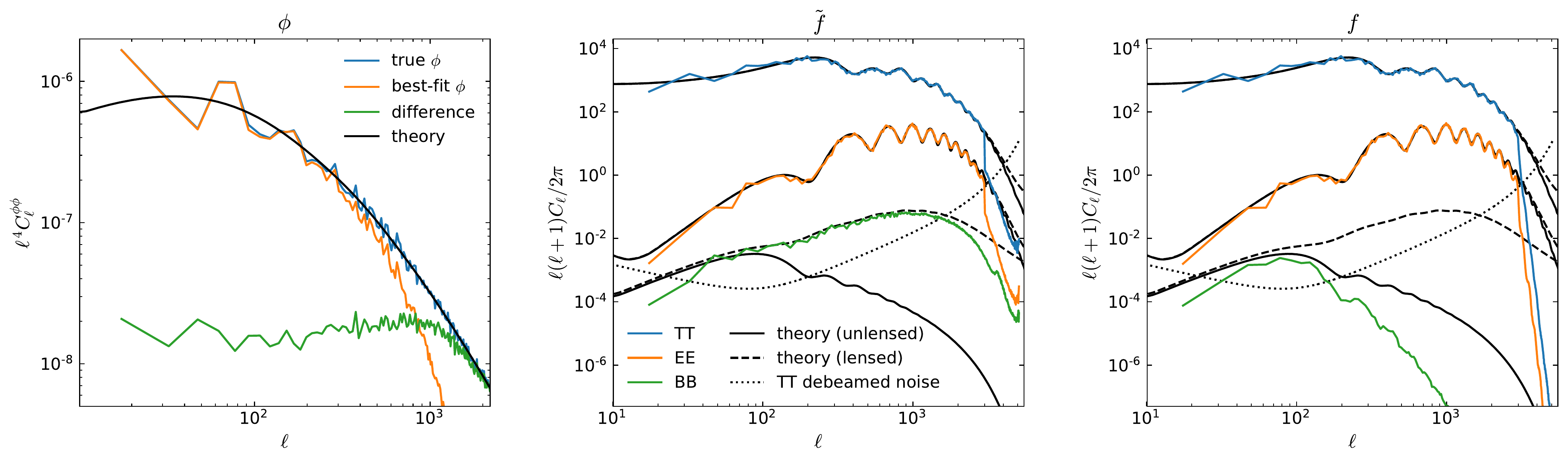}
\end{center}

\caption{The power spectra of the best-fit $\phi$ and lensed/unlensed CMB maps
from a run of our algorithm, as compared to the input theory spectra. This is
for the run with only Fourier-space masking described in
Sec.~\ref{sec:results_nomasking} (the same run for which maps are shown in
Fig.~\ref{fig:nomasking_maps}). The left panel also shows the power spectrum of
the simulation truth for the $\phi$ map itself as well as the power spectrum of
the difference between this and our reconstructed solution, demonstrating the
fidelity of the reconstruction. The ``bump'' visible in the lensed spectra near
the Nyquist frequency at $\ell\,{=}\,3600$ signals the smallest scale for which
the \textsc{LenseFlow} pixelized lensing approximation is accurate at this pixel
size (similar features are produced by other lensing algorithms). We mask the
data in Fourier space beyond $\ell\,{=}\,3000$ so that we are not sensitive to
this region, and the effects of this mask are visible above as a sharp
suppression in power at $\ell>3000$.}

\label{fig:nomasking_spectra} \end{figure*}
%%%%%%%%%%%%%%%%%%%%% END figure %%%%%%%%%%%%%%%%%%%%%

Although Hessians are not needed for our iterating equations, we remark that by
a process analogous to inserting infinitesimal perturbations to
\eqref{eq:deltaflow_ode}, one can create an ODE flow for the lensing Hessian
starting from the Jacobian ODE. This Hessian operator cannot be stored in
practice for realistically sized maps, but can be applied in the same
computational order as the lensing and Jacobian operations themselves. This
could prove very useful for sampling algorithms, for example aiding in computing
the mass matrix in a Hamiltonian Monte-Carlo sampler.

%%%%%%%%%%%%%%%%%%%%% BEGIN figure %%%%%%%%%%%%%%%%%%%%%
\begin{figure}
\begin{center}
\includegraphics[width=\columnwidth]{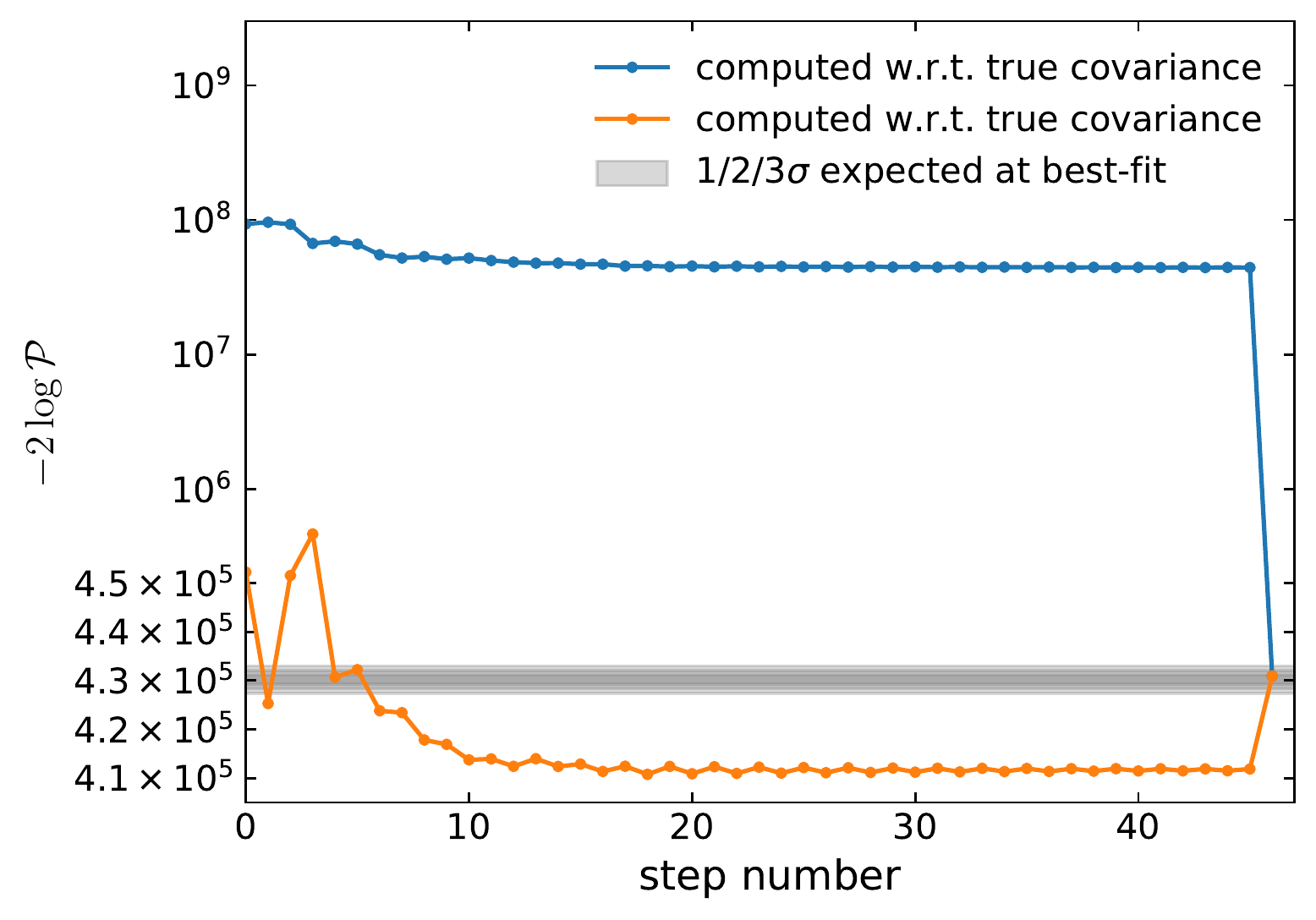}
\end{center}

\caption{The posterior probability after each iteration of our algorithm during
the run on the simulated dataset described in Sec.~\ref{sec:results}. The top
(blue) line is the posterior with respect to the true covariance, and the bottom
(orange) line is with respect to the cooling covariance (note the y-scale is
mixed log and linear). For the final step these two are identical since the
cooling covariance is fully cooled and equals the true covariance. The grey band
represents the value of the posterior probability expected at the best-fit
point, and our best-fit sits well within this expectation.}

\label{fig:lnPtrace}
\end{figure}
%%%%%%%%%%%%%%%%%%%%% END figure %%%%%%%%%%%%%%%%%%%%%

\section{Results}\label{sec:results}

\subsection{Without map-level masking} \label{sec:results_nomasking}

We now begin to test our algorithm on simulatations. We generate simulated data
with CMB-S4 like noise properties, since it is for these low noise levels that
one expects to see a major benefit of the optimal procedure. We assume
$1\,\mu$K-arcmin Gaussian temperature noise, scaled by $\sqrt{2}$ for
polarization, and a 3 arcmin Gaussian beam \citep{abitbol2017}.
Additionally, at low multipoles we adjust the noise power-spectrum to mimic a
$1/f$ knee. Specifically, we take $\ell_{\rm knee}=100$ and $\alpha_{\rm
knee}=3$ according to the parametrization of \cite{barron2017}, who suggest that
for a large aperture array this would be the maximum allowable knee frequency to
be competitive with other configurations. This, in effect, lets us test the
maximal but realistic impact of a non-white noise power-spectrum on our
procedure.

%%%%%%%%%%%%%%%%%%%%% BEGIN figure %%%%%%%%%%%%%%%%%%%%%
\begin{figure*}
\centering
\includegraphics[width=\textwidth]{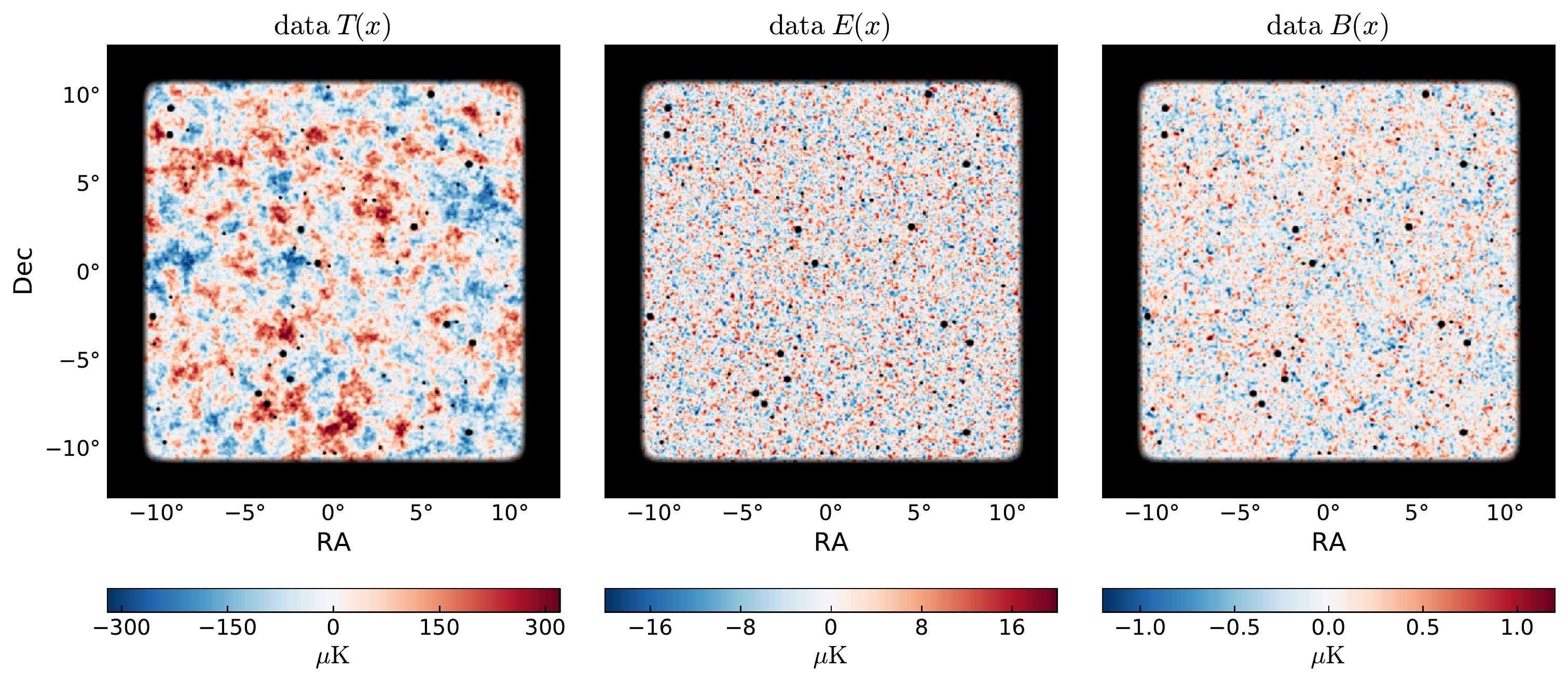}

\caption{The simulated data used in the runs described in
Sec.~\ref{sec:results}. We use a 512$\times$512 grid with 3 arcmin pixels, which
covers roughly 600\,deg$^2$. It assumes a setup approximating an expected CMB-S4
configuration, with a 3 arcmin beam and stationary 1\,$\mu$K-arcmin temperature
noise, modulated to include a $1/f$ contribution below $\ell_{\rm knee}=100$
(see text for more details). One hundred unapodized point sources with radii
between 5 and 10 arcmin are randomly placed within the region. A 2$\degree$
mildly apodized border mask is applied, as well as a Fourier-space cut above
$\ell>3000$. Note that for this figure the mask is simply overlayed on the
unmasked $T$, $E$, and $B$ images rather being multiplied into $T$, $Q$, and $U$
as is done in the likelihood, since multiplying it in would result in large $E$
to $B$ leakage spoiling the ability to see $B$. Additionally, the unmasked data
has been Wiener filtered with the lensed CMB covariance as the signal
covariance to reduce the visual impact of noise.}

\label{fig: data_wmask}
\end{figure*}
%%%%%%%%%%%%%%%%%%%%% END figure %%%%%%%%%%%%%%%%%%%%%

We use pixels which are 3 arcmin on a side, which are fairly large compared to
typical analyses. This highlights one of the advantages of \textsc{LenseFlow},
which is that we get numerically stable and accurate lensing with determinant
equal to exactly unity even on such large pixels. At fixed map size, this makes
the algorithm faster because of the smaller matrix operations involved. The runs
described here use maps which are 512$\times$512 pixels, which at this
resolution correspond to around 600\,deg$^2$, comparable to currently existing
polarization datasets to which our procedure would be naturally applicable
\citep[e.g.][]{story2015,sherwin2016}. The Nyquist frequency for 3 arcmin pixels
is $\ell\,{=}\,3600$, above which we expect little cosmological information in
our setup. Nevertheless, we have also verified the algorithm with 1 arcmin
pixels, and find the main difference is just a longer time-to-convergence for
the conjugate gradient.

We generate a Gaussian random realization of the CMB from a fiducial CMB
spectrum with cosmological parameters given by their posterior mean given the
{\it Planck} 2015 TT data \citep{planckcollaboration2015}, combined with the
updated HFI large scale polarization data $\tau$
\citep{planckcollaboration2016}. We take $r_{0.002}=0.05$, compatible with
current upper bounds \cite{bicep2collaboration2016}.

Using the configuration just described, we create one main simulated dataset.
The resulting temperature and polarization maps are shown in Fig.~\ref{fig:
data_wmask}. Note that although this figure shows a pixel mask, in this section
we consider only Fourier-space masking (we will add map-level masking in
Sec.~\ref{sec:results_masking}). The Fourier mask we use in this section is an
unapodized low-pass filter at $\ell=3000$.

We run 50 iterations of the algorithm on this data, the entire run completing in
around two hours on a single Intel Haswell 2.3GHz 16-core CPU.\footnote{As the
algorithm itself is entirely sequential, no parallelization is employed aside
from using a multi-threaded FFT library and making use of SIMD vectorization for
point-wise matrix multiplications.  The run-time is dominated by computing the
LenseFlow ODE velocity during the Runge-Kutta integration for the lensing
operations performed in the CG step. The asymptotic complexity is set by the FFT
and is thus $O(N\log N)$ where $N$ is the number of pixels in the map, although
in practice we find speed difference between lensing e.g. a 1024$\times$1024 and
2048$\times$2048 map is a bit worse than this because the bottleneck is memory
access.} In Fig.~\ref{fig:nomasking_maps} we see the excellent visual agreement
between the true $\phi$ and lensed and unlensed $B$ maps and the ones recovered
by the algorithm. We expect these should resemble something like a Wiener filter
solution, and thus have low signal-to-noise modes attenuated; the
signal-to-noise is low enough that this is visually apparent only for the
unlensed $B$ map. Fig.~\ref{fig:nomasking_spectra} shows the power-spectrum of
these maps, where one can see the attenuation for all cases, as well as the very
small residual at medium and large scales between the reconstructed $\phi$ map
and the truth.

%%%%%%%%%%%%%%%%%%%%% BEGIN figure %%%%%%%%%%%%%%%%%%%%%
\begin{figure*}
\centering
\includegraphics[width=\textwidth]{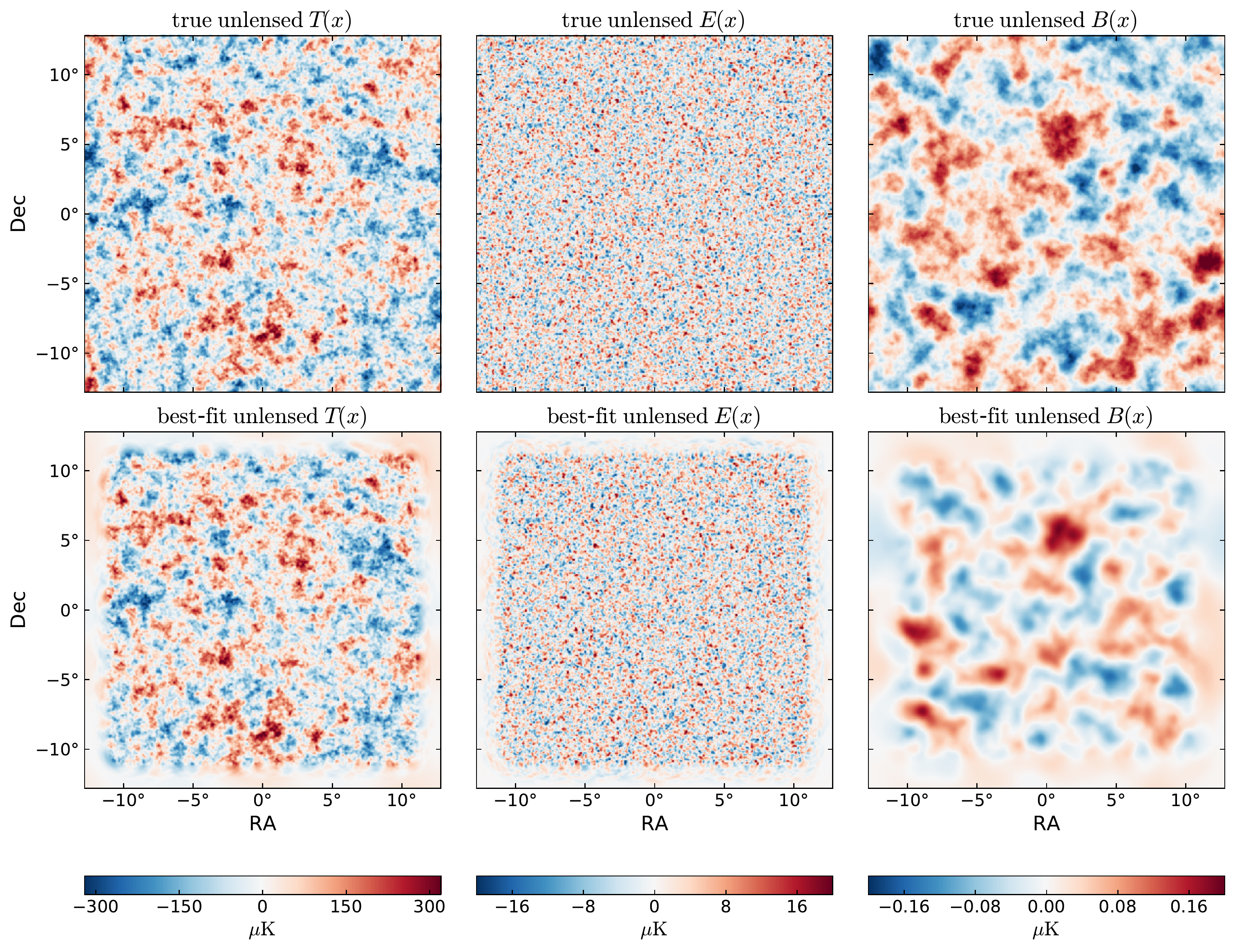}

\caption{The reconstructed unlensed $T$, $E$, and $B$ maps from a run of our
algorithm on simulated data (bottom row), as compared to the simulation truth
(top row). This is for the run discussed in Sec.~\ref{sec:results_masking} which
includes the real-space mask that is visible in Fig.~\ref{fig: data_wmask}. As
expected, low signal-to-noise modes are attenuated and the solution provides a
partial reconstruction even in the masked region.}

\label{fig:masking_maps_TEB}
\end{figure*}
%%%%%%%%%%%%%%%%%%%%% END figure %%%%%%%%%%%%%%%%%%%%%

These maps and power-spectra look as one might expect for a MAP estimate, but we
would like a more robust way to verify that we have attained the true maximum.
One way to do so it to compute the $\chi^2$ expected at the best-fit point and
compare to what we actually achieved. By $\chi^2$, we are referring to the sum
of the terms in \eqref{eq:lnP} excluding the determinants, i.e. the sum of the
$\chi^2$'s of the data residual, $f$, and $\phi$, with respect to $\Cn$,
$\Ccmb$, and $\Clen$, respectively. Approximating the problem as linear, we
expect the best-fit $\chi^2$ to scatter according to a $\chi^2$ distribution
with degrees of freedom given by the total number of umasked pixels in the three
terms, minus the number of free parameters which are fit for. In
Fig.~\ref{fig:lnPtrace} we show the one, two, and three sigma regions for this
expectation as the gray bands. The $\chi^2$ after each of the 50 steps of the
algorithm is also plotted, both with respect to the true covariance, $\Ccmb$,
and with respect to the cooling covariance, $\preCcmb{\hat}$. By the final
iteration when we fully cool the covariance, we are well within this gray band,
a good indication of convergence. 

% We have also run 100 other cases with a
% different realization of the input data. The posterior probability at each of
% their best-fit points are shown as the histogram in the right panel of
% Fig.~\ref{fig:lnPtrace}. These all also fall well within the band, indicating
% the good convergence was not just a fluke of our default run. 

Although this result is suggestive that we have successfully converged,
our problem is not exactly linear, so we cannot rule out that the true expected
distribution of best-fit $\chi^2$ is actually lower. Another test we can perform
is to examine the gradient of the posterior after each iteration. As we reach a
local or global maximum, we expect the gradient to approach zero. Since the
gradient in the $\tilde f$ direction is always reduced to zero up to numerical
precision by the Wiener filter step, we examine the gradient in the $\phi$
direction. Here, we find that across all scales, the power-spectrum of the
gradient drops by several orders of magnitude during the 50 iterations of the
algorithm, until hitting a numerical floor. Taken together, that the best-fit
maps and power-spectrum look as expected given the simulation ground truth, that
we are close to the expected $\chi^2$, and that the gradient is approaching zero
are strong indications that the algorithm has reached the global maximum.

\subsection{With map-level masking} \label{sec:results_masking}

We now turn to demonstrating that the algorithm works when we apply map-level
masking. Such masking is necessary in any real analysis as various sources of
galactic and extragalactic contamination are most efficiently dealt with by
directly excising them from the maps. Here we randomly place 100 point sources
holes with radii between 5 and 10 arcmin. Additionally, for a flat-sky analysis
as performed here, it is necessary to include a border mask so as to ``embed''
the observed sky patch (which is non-periodic) onto a Fourier grid with is
otherwise assumed periodic. To this end, we apply a 2\degree\ border mask. Both
the border mask and the point source mask are mildly apodized. 

We use the identical simulated data shown in Fig.~\ref{fig: data_wmask} as in
the previous section, with the only change being that we apply this map-level
mask. Note that we continue to apply the Fourier mask which removes $\ell>3000$,
hence here we are testing the performance of the algorithm in the presence of
masking which is not diagonal in either map or Fourier space. This introduces a
subtle non-triviality in inverting the noise covariance of the masked data,
which we account for here with a trick of filling in the masked regions of the
map with a realization of noise from $\Cn$. The data, as well as the mask, is
shown in Fig.~\ref{fig: data_wmask}. 

Two small changes to the algorithm itself are necessary as compared to the
unmasked run. First, the cooling weights are recomputed for the specific mask,
although using the same procedure as described earlier. Second, not
surprisingly, the Wiener filter requires more steps to achieve satisfactory
accuracy.\footnote{In fact, to ease convergence in some cases we find it
necessary to replace the one-dimensional line-search $\phi_i - \alpha\Clen g$
over $\alpha$ with a two dimensional line-search $\phi_i - \alpha_1\Clen
g-\alpha_2\psi$ over $(\alpha_1, \alpha_2)$ where $\psi$ is defined as the
inverse Laplacian of the border mask and is designed to approximate the
mean-field feature described later in Section \ref{sec:results_masking}. This
modification appears to improve numerical stability in Algorithm \ref{maxalg},
but is not necessary in all configurations we have tried, so we mention it here
but do not discuss it further.} That no other major changes to the algorithm are
required might have been expected because, as mentioned earlier, one
fundamentally nice feature of the lensed parametrization is that it removes from
the $\phi$ step any explicit dependence on the instrument or dataset (i.e. on
masking). Of course, there could have been an impact on the decorrelating effect
of switching to the lensed parametrization itself, or on the effectiveness of
the quasi Newton-Raphson step, but neither appears to be the case. This is good
news as it means that if one wishes to even further improve the performance of
the algorithm, one needs to focus only on improving the Wiener filter, where
many more sophisticated methods exist other than the fairly rudimentary
preconditioned conjugate gradient which we have found sufficient here
\cite[e.g.][]{smith2007,elsner2013,seljebotn2014,huffenberger2017,kodiramanah2017}.q

Fig.~\ref{fig:masking_maps_TEB} shows the unlensed CMB estimate $\hat
f_J$ compared the simulation truth. We find, as expected, a Wiener filter-like
solution with low signal-to-noise modes attenuated as is visible for $B$, and
with power slowly decaying towards zero in the masked regions as is visible for
$T$, $E$, and $B$. 

\begin{figure*}
\centering
\includegraphics[width=\textwidth]{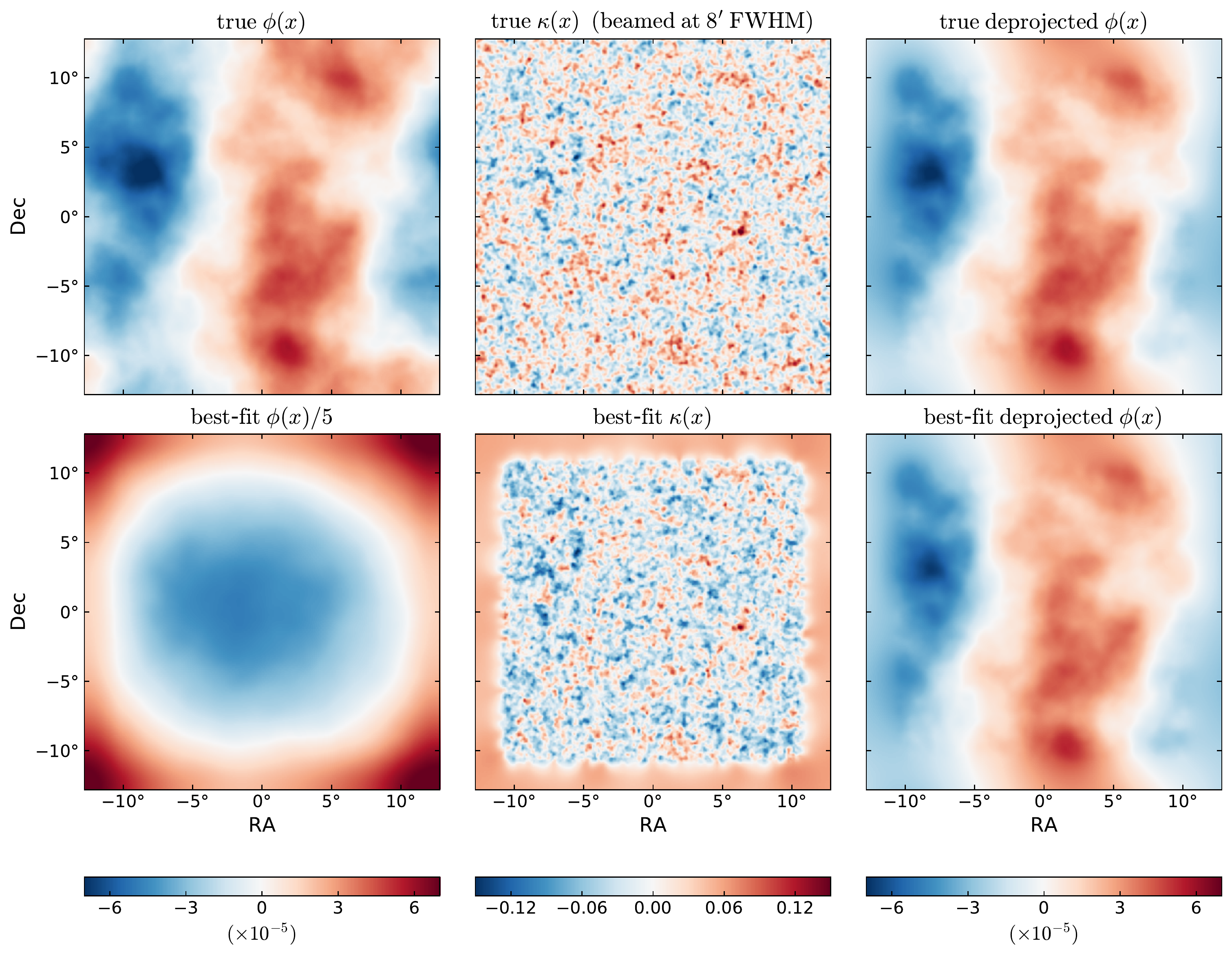}

\caption{The reconstructed lensing potential from a run of our algorithm on
simulated data (bottom row), as compared to the simulation truth (top row). The
first column is the raw $\phi(x)$ map that maximizes the posterior. The middle
column is the corresponding convergence, $\kappa(x)\equiv-\nabla^2\phi(x)/2$,
which allows one to see the good agreement with the truth in the unmasked
regions. A small uniform negative ``mean-field'' correction inside the mask is
visually recognizable as a slight preponderance of blue. The final column is
after deprojecting this mean field using the procedure described in
Sec.~\ref{sec:results_masking}, allowing one to better recognize the agreement
with the true $\phi$ map.}

\label{fig:masking_maps_phi}
\end{figure*}

The lensing potential estimate $\hat \phi_J$ corresponding to $\hat f_J$ is
shown in Fig.~\ref{fig:masking_maps_phi} (bottom left). Notice what appears to
be a large scale ``bias'' in the estimate $\hat \phi_J$ as compared to the true
$\phi$ (top left). This feature corresponds to a so called ``mean field'', akin
to the one which must be subtracted to debias the quadratic estimator. Similarly
as for the quadratic estimate, it arises because the mask induces correlations
between different $\ell$-modes, which the best-fit then attributes to lensing.
We remark that the marginal estimate $\hat\phi_M$ would not show this feature
because it is implicitly corrected for by the determinant term found in the
marginalized posterior \eqref{eq:laplace_approx} which is not present in the
joint posterior \eqref{eq:lnP}. 

The effect of the mean field bias in $\hat\phi_J$ is simpler when considering
the convergence $\kappa \equiv -\nabla^2 \phi/2$. There, the mean field roughly
translates to an additive constant offset over non-masked pixels,
\begin{align}
    \text{ $\hat\kappa_J(x) \approx \mu + \kappa(x)$ \;for all non-masked pixels $x$}.
\end{align}
Intuitively this can be understood as follows. Because in the masked regions the
Wiener-filter like suppression drives the solution to zero, in the absence of
lensing this leads to an $f$ power spectrum which, on average across the entire
map, is smaller than expected given $\mathcal C_f$. Now note that since the CMB
has a mostly ``red'' spectrum (i.e. tilted to the right), an overall
magnification has a similar effect to reducing the overall
amplitude.\footnote{This degeneracy is in fact exact for power-law spectra in
the limit of infinite-size maps \citep{anderes2010}.} Thus with the lensing
potential available as a free parameter, the best-fit is able to slightly
increase $f$ to better agree with with its covariance, but add an overall
magnification to $\phi$ so that $\tilde f$ is reduced and still agrees with the
data. 

\begin{figure}
\centering
\includegraphics[width=\columnwidth]{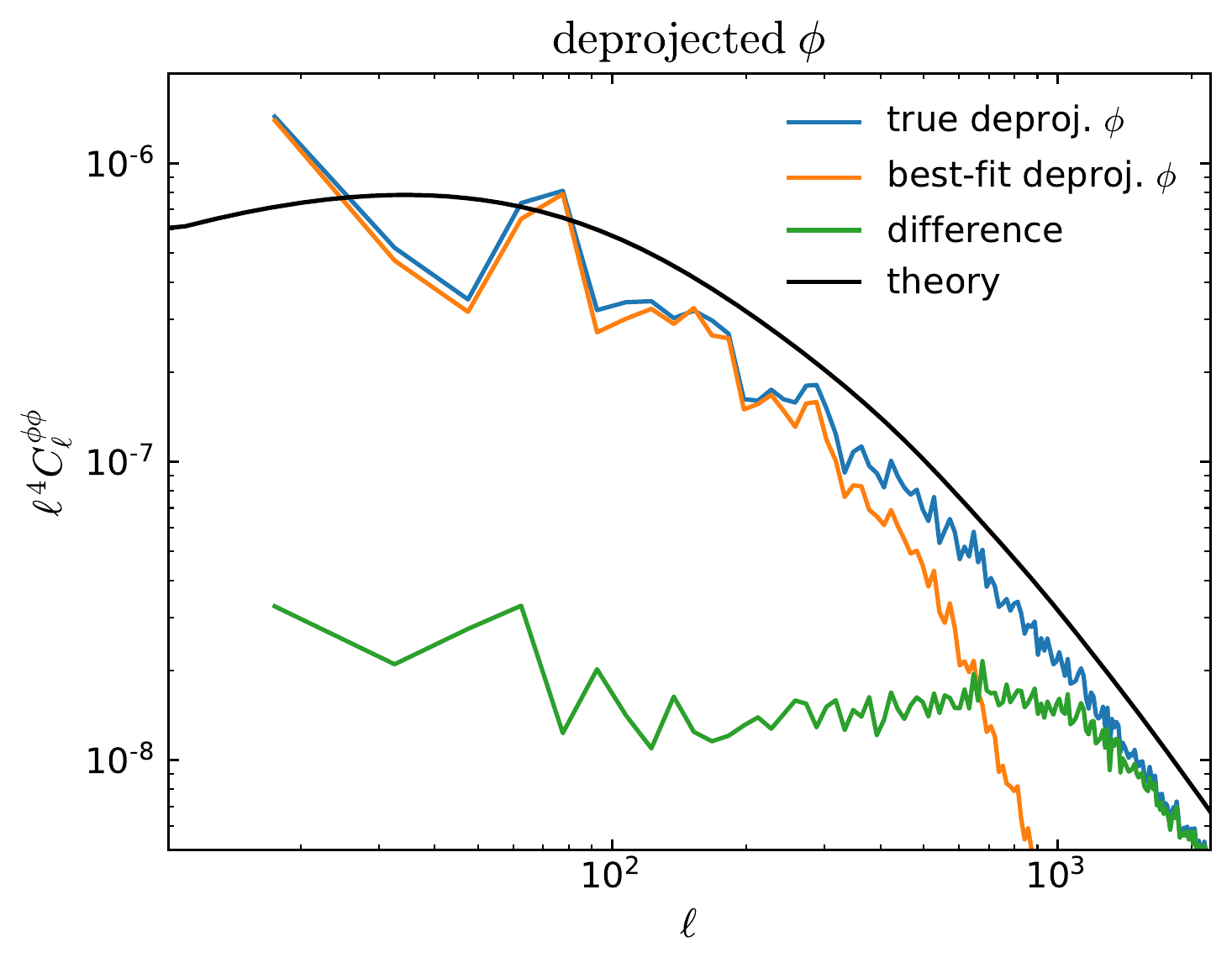}
\caption{The power spectra of the best-fit $\phi$ map as compared to the
simulation truth and theory spectrum for the run with real-space masking
described in Sec.~\ref{sec:results_masking}. The best-fit and simulation truth
$\phi$ maps are the ones shown in the right column of
Fig.~\ref{fig:masking_maps_phi} and have had the mean-field deprojected
according to the procedure described in Sec.~\ref{sec:results_masking}.}
\label{fig:masking_spectra_phi} \end{figure}

This effect can be seen in the middle column of Fig.~\ref{fig:masking_maps_phi}
where the fluctuations of $\hat\kappa_J(x)$ (bottom middle) track the true
$\kappa(x)$ (top middle, plotted with an additional beam to make the relevant
scales more visible). Notice that the average value of $\hat\kappa_J(x)$ over
non-masked pixels appears slightly smaller than zero. This is the mean field and
results in a more visually dramatic effect on the original non-Laplacian scale
(as seen in the bottom left image). To probe the accuracy of the smaller scale
fluctuations one can re-center $\hat\kappa_J$ and  $\kappa$ to have zero mean
over non-masked pixels, then set any masked pixels to zero so that only errors
within the observation region are probed. The resulting error bandpowers are
shown in Fig.~\ref{fig:masking_spectra_phi} and can be seen to be similar to
what one expects from non-masked observations. Applying $-2\nabla^{-2}$ to the
re-centered and mask-attenuated $\hat\kappa_J$, which we refer to as
``deprojecting'' in the figure captions, has the effect of visually removing the
mean field features in the original estimate (shown bottom right in
Fig.~\ref{fig:masking_maps_phi} with the corresponding operation applied to the
true $\phi$ shown top right).

As in the previous section, we would like to confirm convergence, thus
ascertaining that the mean-field is a real feature of the global MAP estimate
and not a local mode or artifact of Algorithm \ref{maxalg}. The first piece of
evidence is that the best-fit, similarly as before, attains an acceptable
best-fit $\chi^2$, in this case $0.8\,\sigma$ above expectation. Going beyond
just this one simulated dataset, we also check the distribution of best-fit
$\chi^2$'s on 100 other simulations (with somewhat smaller map sizes for speed
but still with a border mask). The best-fit $\hat \phi_J$ for each of these
displays a qualitatively similar mean-field, while their best-fit $\chi^2$
appear to be in line with expectation as shown in Fig.~\ref{fig: ensemble of MF
simulations}. Finally, we check that even initializing Algorithm \ref{maxalg} at
the true $\phi$ results in the same mean field feature in $\hat \phi_J$ and a
similar best-fit $\chi^2$ value.

As the final piece of evidence that the mean field is a necessary feature of the
joint MAP estimate of $\phi$, we show that similar biases occur naturally in
other MAP estimates for models which have more parameters than data and thus
yield highly non-Gaussian posteriors. Consider the following toy example which
is relevant to the problem of estimating scalar-to-tensor ratio $r$ and which
will foreshadow the discussion in the next section where we free $r$ as a
parameter. 

Suppose we observe a noisy signal which is the product of some scaling
parameter, $r$, with some Gaussian random field, $B$, 
\begin{align}
    d = r B + n
\end{align}
where $n$ is stationary noise and $n$ and $B$ have known spectral densities
$\mathcal C_n$ and $\mathcal C_B$, respectively. Notice that for a given value
of $r$, the maximum of $B \mapsto \mathcal P(r, B \,|\, d)$ is given by a
Wiener filter-like solution,
\begin{align}
    \hat B(r) \equiv r\mathcal C_B (\mathcal C_n + r^2 \mathcal C_B)^{-1} d
\end{align}
Therefore, the joint MAP estimate of $r$ and $B$ can be computed by maximizing
$r \mapsto \mathcal P(r, \hat B(r) \,|\, d)$. However, a simple calculation
shows that this function is always maximized at $r\,{=}\,0$.\footnote{This
statement depends on the prior one takes on $r$, e.g. the singularity is at
$r\,{=}\,0$ with a Jeffrey's prior as we have assumed here, but at
$r\,{=}\,\infty$ with a flat prior. Nevertheless, no reasonable data-independent
prior can remove the singularity entirely, which is the important part of our
example.} The cause of this singularity is simply that there is a perfect
degeneracy in the likelihood term wherein one can decrease $B$ and increase $r$
and fit the data identically. The best-fit of the full posterior will then
maximize just the prior along this slice of parameter space, which in this case
happens at $r\,{=}\,0$. Yet, the posterior expected value of $r$, which
effectively marginalizes over the unknown $B$, gives a perfectly normal and
non-zero estimate of $r$. To complete the analogy, note the similarity in data
residual between the lensing case and our toy example, $d - \mathcal L(\phi)f$
and $d - rB$. Thus, for similar reasons as in this toy example, the MAP estimate
of $\hat \phi_J$ is driven away from its expected value, although due to the
non-perfect degeneracy we are not driven all the way to any singularities at
zero. 

Our point with this example is to demonstrate that MAP estimates need not be
optimal, and to stress that while MAP estimates can have poor properties as
estimators (such as in this case for $r$), sampling the posterior will always
yield the correct answer. Nevertheless, the fact that $\hat \kappa_J$ tracks
fluctuations of $\kappa$ with little apparent bias suggests $\hat \kappa_J$
could still form a useful estimator, and moreover potentially be more useful for
initializing a sampling algorithm for the joint posterior.

%%%%%%%%%%%%%%%%%%%%% BEGIN figure %%%%%%%%%%%%%%%%%%%%%
%%%%%%%%%%%%%%%%%%%%% END figure %%%%%%%%%%%%%%%%%%%%%

%%%%%%%%%%%%%%%%%%%%% BEGIN figure %%%%%%%%%%%%%%%%%%%%%
%%%%%%%%%%%%%%%%%%%%% END figure %%%%%%%%%%%%%%%%%%%%%

%%%%%%%%%%%%%%%%%%%%% BEGIN figure %%%%%%%%%%%%%%%%%%%%%
\begin{figure}
\centering
\includegraphics[width=\columnwidth]{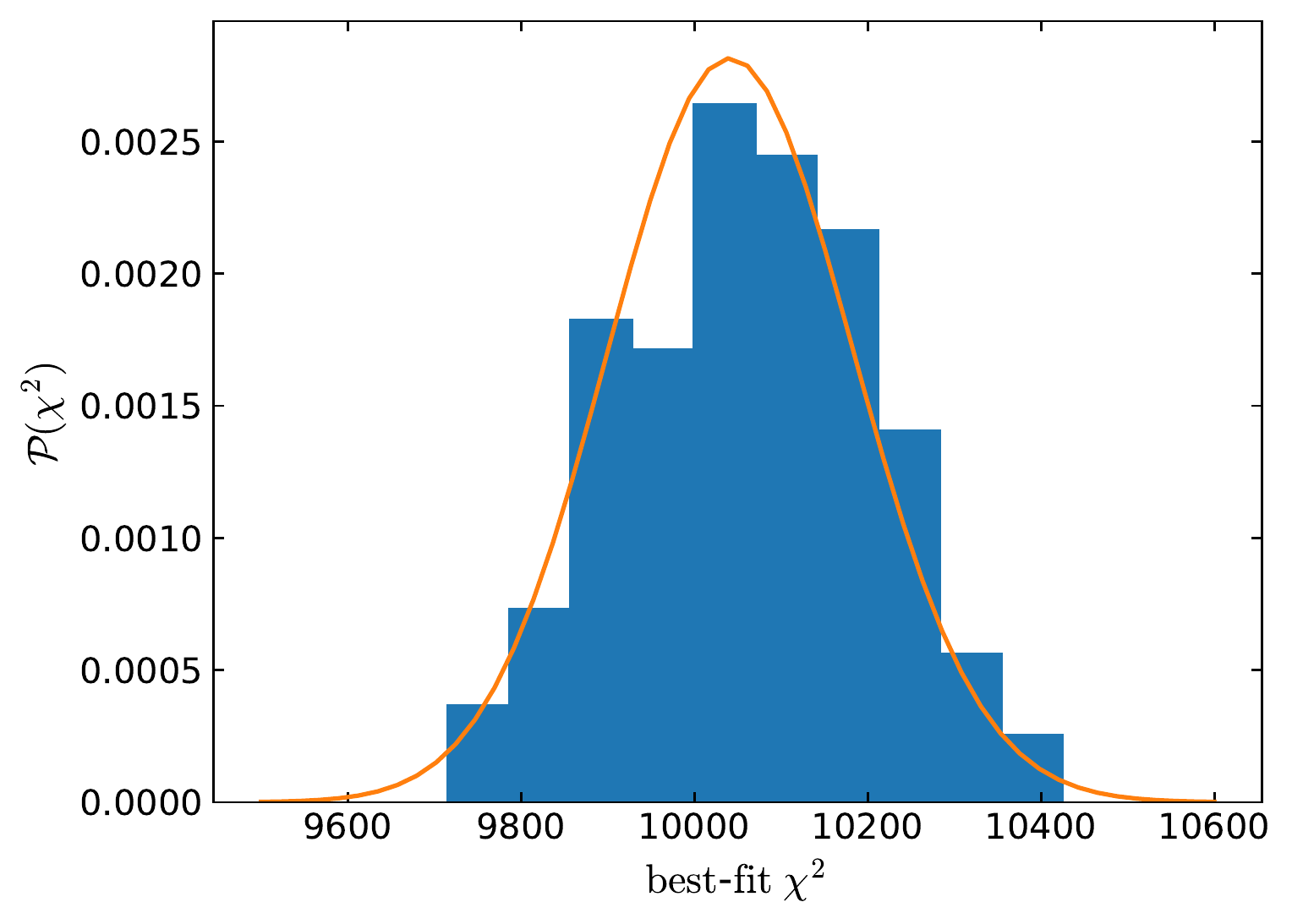}

\caption{Distribution of the $\chi^2$ of the best-fit point from runs on 500
different simulated datasets. For speed, we have reduced the map size as
compared to the main runs described in this work to 128$\times$128 pixels (while
keeping the relative width of the border mask width) and use only $E$ and $B$.
The expected distribution of the best-fit $\chi^2$ under a Gaussian
approximation of the posterior is shown as the orange curve.}

\label{fig: ensemble of MF simulations}
\end{figure}

% \pagebreak
\subsection{With $r$ as a free parameter} \label{sec:results_r}

\begin{figure}
\begin{center}
\includegraphics[width=\columnwidth]{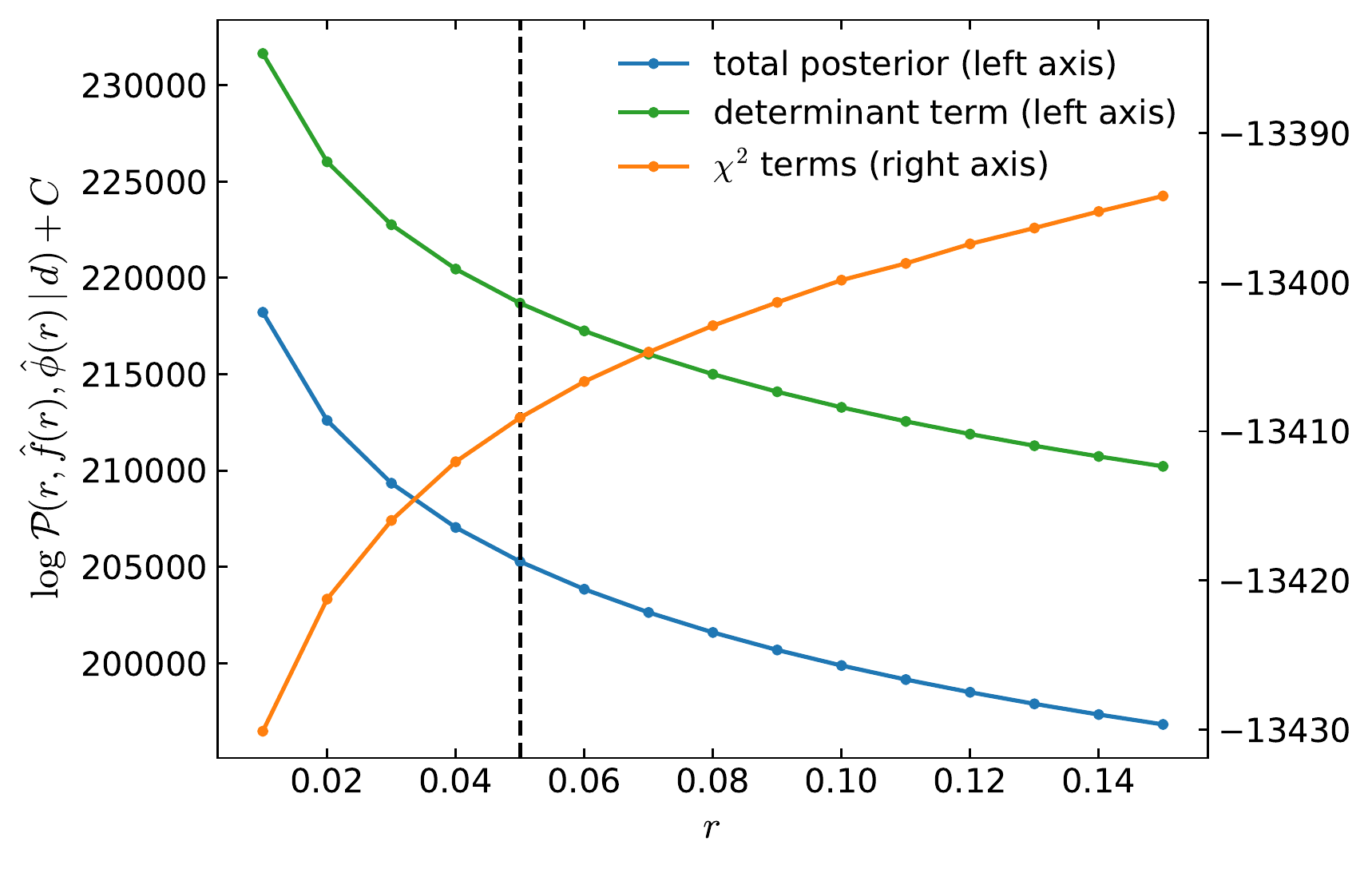}
\end{center}

\caption{A slice through the joint posterior probability \eqref{eq:lnP}, varying
$r$ and maximizing with respect to $f$ and $\phi$ for each value of $r$. For
speed, we have reduced the map size as compared to the main runs described in
this work to 128$\times$128 pixels (while keeping the relative width of the
border mask the same). The green curve (left axis) is the contribution from
$\det \Ccmb(\theta)$, the orange curve (right axis) is the contribution from the
three $\chi^2$ terms (i.e. the first three terms of \eqref{eq:lnP}), and the
blue (left axis) is the sum of these two. This demonstrates that the joint MAP
estimate of $r$ is not useful as it is driven to zero. The lack of apparent
numerical noise in the orange curve demonstrates the stability of the
maximization algorithm.}

\label{fig:restim}
\end{figure}

The toy example from the previous section serves a dual purpose, as it was
selected to prepare discussion of the actual problem of $r$ estimation. The
differences are that in reality we have tensor contributions to $T$ and $E$ in
addition to just $B$, and of course because the toy example did not involve
lensing. Nevertheless, we might expect qualitatively similar behavior, and in
this section we verify that this is indeed the case. 

To do so, we generate simulated data with $r\,{=}\,0.05$ then run the
maximization algorithm for $\mathcal P(f,\phi\,|\,d,r)$ over a grid of $r$
values from $r\,{=}\,0$ to $r\,{=}\,0.15$. More specifically, we compute,
\begin{align}
    \big(\hat f(r), \hat \phi(r)\big) = \argmax_{f,\phi} \;\mathcal P(f,\phi\,|\,d,r)
\end{align}
and plot the function $r \mapsto \mathcal P(r,\hat f(r),\hat \phi(r)|\,d)$ as
the blue curve in Fig.~\ref{fig:restim}. Indeed we find that a singularity at
zero exists, which confirms that the MAP estimate of $r$ (jointly with $f$ and
$\phi$) is not a useful estimator, as it is always zero.  

We point out that the total posterior plotted in blue is largely dominated by
just the determinant of the CMB covariance in \eqref{eq:lnP}, $\det \Ccmb(r)$.
This is independent of $f$ and $\phi$ and hence independent of the maximization
algorithm; to see the performance of the maximization, we plot in orange the
contribution to the total posterior from only $\chi^2$ terms, i.e. the first
three terms of \eqref{eq:lnP}. The smoothness of this curve is further evidence
of the quality of convergence, as we might otherwise expect to see lots of
numerical noise in adjacent bins. 

This convergence is important because the orange curve gives one contribution to
the full marginal posterior, $\mathcal{P}(r)$, and if this piece were not stable
numerically, adding in the other contributions would be of no use. Indeed, under
the Laplace approximation we can compute the marginal posterior by just adding
in a determinant term, i.e. the analog of the denominator in
\eqref{eq:laplace_approx} but for marginalization over {\it both} $f$ and
$\phi$, and which would cancel out the singularity seen here. In fact, something
like this could potentially be calculable in practice with Hessian operators and
if one can compute accurately enough the necessary determinant via Monte-Carlo.
Ultimately, we seek to sample directly from the exact posterior, producing a
marginal $\mathcal{P}(r)$ with no approximation. Again, the stability of the
curves in Fig.~\ref{fig:restim} suggest this should be numerically possible as
long as satisfactory convergence of the sampling algorithm can be achieved.

\section{Lensing determinant} \label{sec:lensdet}

We now revisit in more detail a discussion surrounding the lensing determinant.
One key point which is worth stating explicitly is that, in the limit of
infinite resolution maps, the lensing operation is unique. It is only upon
considering pixelixed maps (where one necessarily looses some information) that
there is any room for different lensing algorithms to exist. Indeed, a number of
such algorithms have been given in the literature \citep{lewis2005,
hamimeche2008, lavaux2010, louis2013}, all of which asymptote to ``true''
lensing in the limit of infinitely small pixels, but on discrete maps differ in
how they reconstruct the information lost due to pixelization.

With infinite resolution, or equivalently with band-limited maps, the
determinant of lensing is unity. This follows directly from our proof that
\textsc{LenseFlow} gives a lensing determinant of exactly one in the limit of
continuous integration regardless of spatial resolution
\eqref{eq:lenseflow_det}, and therefore also for infinite resolution. This proof
applies as long as the matrix $M_t$ is invertible, i.e. in the weak-lensing
regime.

An important question is then, given typical CMB spectra and pixelizations, are
our maps close enough to band-limited that might be able to use any generic
lensing algorithm and ignore the determinant? To that end, we perform the
following test.

%%%%%%%%%%%%%%%%%%%%% BEGIN figure %%%%%%%%%%%%%%%%%%%%%
\begin{figure*}
\begin{center}
\includegraphics[width=\textwidth]{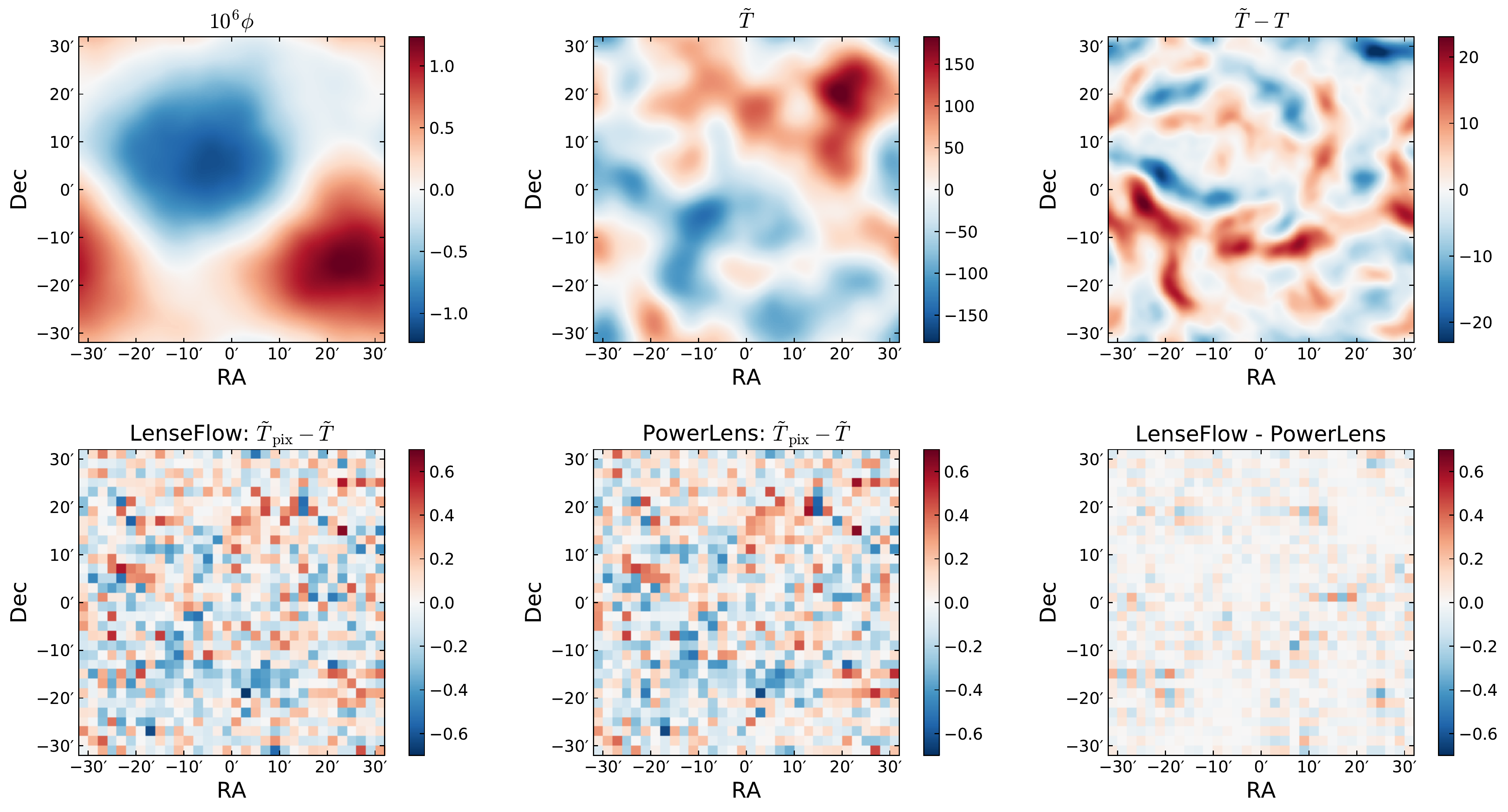}
\end{center}

\caption{An example of the difference between \textsc{LenseFlow} and
\textsc{PowerLens} pixelized lensing algorithms. The top row shows a simulated
lensing potential $\phi$, a simulated temperature field lensed by this potential
$\tilde T$, and the difference between the lensed and unlensed field, $\tilde T -
T$. Here we have used a high resolution (1/8\,arcmin) pixelization such that
lensing is essentially exact on relevant physical scales, independent of
algorithm. In the bottom row, we compare this to pixelized lensing for the exact
same patch of sky and same simulated $\phi$ and $T$. That is to say, we first
pixelize $\phi$ and $T$ to a coarser resolution (2\,arcmin), then apply lensing,
then compare to a pixelized version of the true lensed field from the top row.
The first two panels show the result for \textsc{LenseFlow} and
\textsc{PowerLens}, and the third is the difference between the two. We stress
that the features in the final panel are not numerical artifacts, they represent
real differences between how the two algorithms extrapolate sub pixel-scale
fluctuations. It is exactly these differences that give rise to the different
determinants for the two lensing operations.}

\label{fig:lenseflow_pixlens}
\end{figure*}
%%%%%%%%%%%%%%%%%%%%% END figure %%%%%%%%%%%%%%%%%%%%%

%%%%%%%%%%%%%%%%%%%%% BEGIN figure %%%%%%%%%%%%%%%%%%%%%
\begin{figure}
\includegraphics[width=\columnwidth]{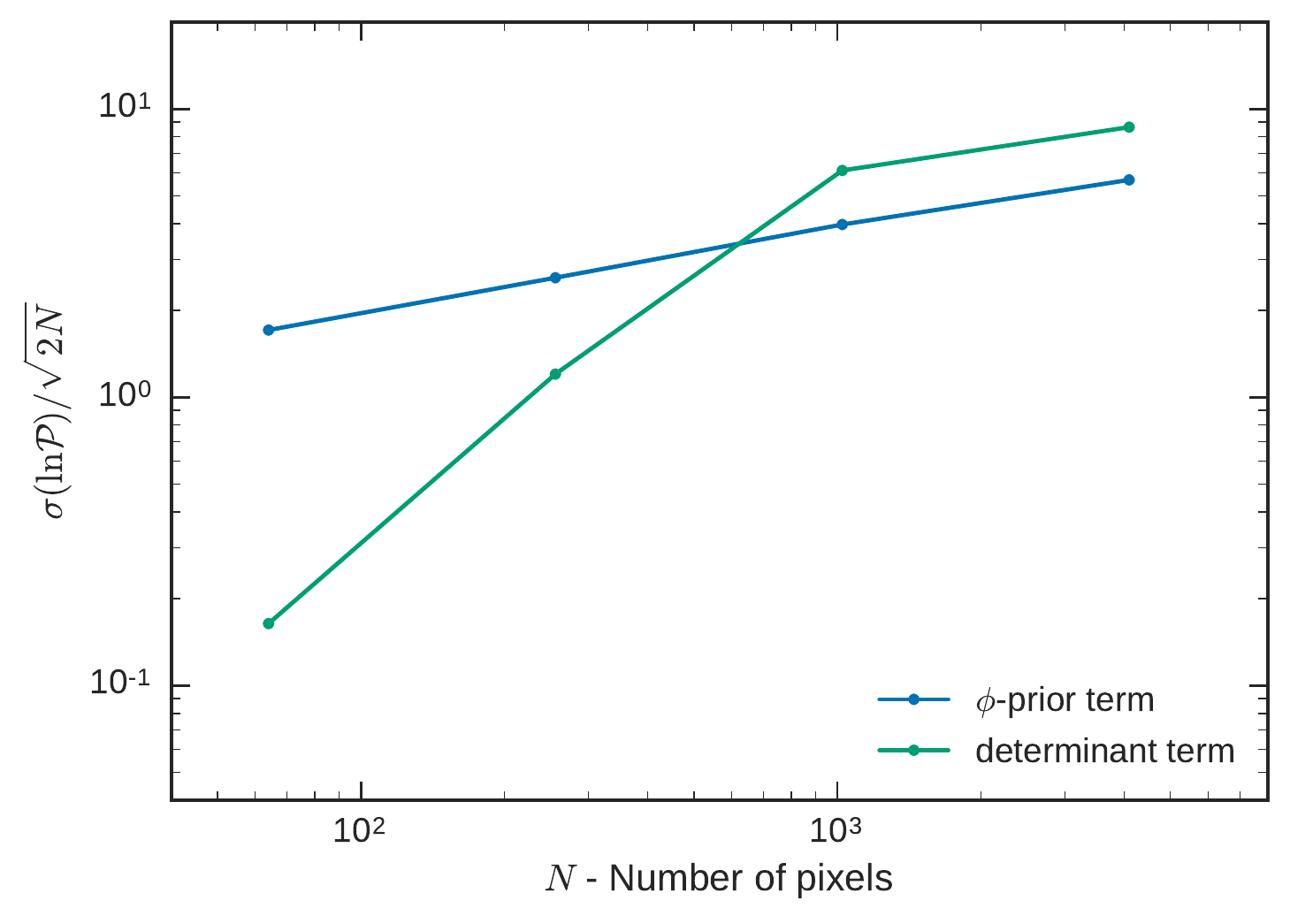}

\caption{The standard deviation of the variation in the log probability values
for the $\phi$-prior term, $\phi^\dagger \Clen^{-1} \phi$, and lensing
determinant term, $2\log \left|\det \Lphi{\phi}\right|$, in \eqref{eq:lnPlensed}, as
computed from Monte-Carlo samples of $\phi$. These samples approximate samples
from the posterior probability $\mathcal{P}(f,\phi\,|\,d)$ for some simulated
data, $d$, assuming full-sky temperature-only reconstruction noise. Here we have
used 7th order Taylor series lensing on 1 arcmin pixels with temperature-only
data. Because the variation in the two terms is of similar order, the
determinant cannot be ignored.}

\label{fig:lnP_terms}
\end{figure}
%%%%%%%%%%%%%%%%%%%%% END figure %%%%%%%%%%%%%%%%%%%%%

For relatively small numbers of pixels, it is computationally feasible to check
by explicitly calculating the matrix representation of $\Lphi{\phi} $ for a
given $\phi$ and taking its determinant.\footnote{This can be done by applying
the operator to some set of maps which form a complete basis. It may also be
possible to use other methods to compute the determinant, we have chosen this
route only for simplicity.} We have done so for map sizes between $8\times8$ and
$64\times64$, and for the standard approximation to lensing where one expands in
a Taylor series around the deflection,
\begin{align}
    \tilde f(x) = f(x+\nabla \phi(x)) = f(x) + \nabla^i \phi(x) \nabla^i f(x) + ...
\end{align}
To check whether one can simply use fairly small pixels, we have performed the
test here with 1 arcmin pixels, i.e. somewhat smaller than the 3 arcmin
pixels we use in the rest of this paper. For this pixel size, the determinant of
the Taylor series lensing approximation asymptotes by the 7th order term in the
expansion. By using this many terms, we are testing the determinant due to the
implicitly assumed sub-pixel extrapolation method of the Taylor series
expansion, rather than the determinant due to Taylor series truncation error.

The exact value of the determinant is, in fact, unimportant; instead, what is
important is how it varies as a function of $\phi$ near the peak of the
probability distribution as compared to the other terms in the posterior
probability. As a simple way to mimick samples of $\phi$ near this peak, we
approximate the problem as a Wiener filter problem, and use the analytic
calculation of the effective reconstruction noise, $\mathcal N_\phi$, from the
iterated full sky quadratic estimator \citep{smith2012}. We expect the
determinant will be most important when the effective noise is high, such as
when performing a temperature-only reconstruction; since we want our method to
work for these cases, we check using the temperature-only $\mathcal N_\phi$.
Finally, we have not upscaled the reconstruction noise for our smaller $f_{\rm
sky}$, thus this check will represent a lower bound on how important the
determinant might be. To mimick the samples of $\phi$, we first simulate a one
single typical best-fit (i.e. ``Wiener filtered'') $\phi$, which is given from
the covariance $\Clen (\Clen + \mathcal N_\phi)^{-1} \Clen$. We then simulate
many samples from around the peak which are given by an additive contribution
drawn from $\Clen (\Clen + \mathcal N_\phi)^{-1} \mathcal N_\phi$. For each of
these samples, we calculate the prior and lensing determinant terms in
\eqref{eq:lnPlensed}. We consider the scatter in the prior term a proxy for the
level of change we might be able to tolerate, and this should be a fairly good
proxy since this term dominates the posterior at the smallest scales to which we
expect the determinant to be most sensitive. Fig.~\ref{fig:lnP_terms} shows the
results. We find that the determinant term varies roughly on the same order as
the prior term, even sometimes larger. Hence it does not appear that it can be
ignored, at least not on the scales probed by these maps (which are, indeed,
relevant physical scales in general).

\section{Conclusions and future work}

In this work, we have presented the first algorithm which produces the joint MAP
estimate of $\phi$, $f$ and cosmological parameters like $r$. There are two
important aspects to the algorithm. First, a change of variables from the
unlensed field, $f$, to the lensed one, $\tilde f$, greatly reduces the
correlations in the posterior making maximization work much more efficiently.
Second, the maximization is a coordinate descent over $\tilde f$ and $\phi$,
which breaks the problem into two clean pieces, one a robustly solvable Wiener
filter problem and the other entirely independent of the instrument and data. 

The workability of the algorithm depends on using a new lensing algorithm which
we have developed called \textsc{LenseFlow}, which has determinant equal to
unity, and allows us to trivially perform the aforementioned change of
variables. While true lensing (i.e. lensing in the limit of infinite resolution)
has determinant equal to unity, \textsc{LenseFlow} appears unique amongst known
algorithms in preserving this property on pixelized maps; although we have only
explicitly verified the determinant for the Taylor series approximation, it
seems unlikely that other algorithms would have this property without it having
been constructed intentionally. Nevertheless, it is worth checking other
algorithms as perhaps their determinant is close enough to unity that it can be
ignored, in which case there could be benefits of speed or convenience to using
them instead. For example, the current implementation of \textsc{LenseFlow} is
likely prohibitively slow on the full sky, and we leave the solution of this
problem to another work. 

Independently of how we have used it here, \textsc{LenseFlow} is interesting
theoretically as a new formulation of lensing. To date, it has clearly been a
very useful tool for cosmologists to work with the Taylor series expansion for
weak lensing; we would argue that the ODE expansion presented here should be a
valuable addition to any cosmologists' ``toolbox'' as well, as it can in some
cases be quite advantageous to work with. For example, we have used it to give a
simple proof of the area-preserving nature of true lensing. Additionally, it is
very convenient that inverses and adjoints are so easily calculated with
\textsc{LenseFlow}, not just for lensing but also for the Jacobian and Hessian
operators. Some of these are possible to calculate with other identities
\citep[e.g.][]{carron2017}, but the \textsc{LenseFlow} solution is very
straight-forward conceptually.

We have also discussed the relationship between the joint posterior,
$\mathcal{P}(f,\phi\,|\,d,r)$, and the marginal posterior,
$\mathcal{P}(\phi\,|\,d,r)$, the latter which is the basis of the algorithm
given by \cite{carron2017}. It is important to note that neither MAP estimate,
$\hat \phi_M$ nor $\hat \phi_J$ (as defined in \eqref{eq: phi_M def} and
\eqref{eq: phi_J def}), is truly optimal in the least-squared sense. The optimal
estimate is $\langle \phi \rangle$ which differs from both due to the
non-Gaussianity of the posterior. The two estimates $\hat\phi_J$ and
$\hat\phi_M$ differ from each other by a mean-field correction, as do the
corresponding delensed estimates $\hat f_J$ and $\hat f_M$, and we have
elucidated the relation between all of these quantities in the context of a
Laplacian integration. One is free to take any of these quantities as an
estimator and debias and quantify its uncertainties via simulations, and this
would certainly lead to improvement over the quadratic estimate. However, any
such procedure would suffer from the problem of needing to assume a value for
$r$ for these simulations, and perhaps from requiring too large a computational
cost, so it is unclear if that is the right way to proceed forward.

Instead, the goal in Bayesian parameter inference is to quantify uncertainties,
e.g. by obtaining samples from the posterior via Markov-Chain Monte-Carlo
techniques. To be efficient, any such sampling algorithm likely needs to
evaluate the gradient of the posterior at each sampled point. When sampling the
marginalized posterior $\mathcal{P}(\phi,r\,|\,d)$, this gradient has a
contribution from a determinant term which is computationally costly as it must
be computed by averaging over simulations. Conversely, the joint posterior
$\mathcal{P}(f,\phi,r\,|\,d)$ does not have such a determinant, and can thus be
sampled from much faster. Once the Markov-Chain is burned in, as long as the
correlation length for $\phi$ is less than the number of simulations needed for
the determinant calculation in the marginalized case, sampling in the joint case
is faster. As shown by \cite{carron2017}, around 500 such simulations are
needed, so there is potential for a large speed-up. Additionally, sampling in
the joint case could make use of exact Hessians of the posterior computed with
\textsc{LenseFlow}, but it is unclear if the Hessian of the determinant that
appears in the marginalized case is calculable.

Of course, if one is interested in posterior samples of the field $f$ itself as
a data product, then one must necessarily sample the joint probability function.
The tools we have developed in this work move us significantly closer to this
goal, and may be useful in their own right in other contexts. 

\par
\ 
\begin{acknowledgements} 
\small
We thank Antony Lewis and Thibaut Louis for helpful discussions during the
course of this work.  EA is supported in part by NSF CAREER DMS-1252795 and a
University of California Davis Chancellor's Fellowship. MM and BDW are supported
by the Labex ILP (reference ANR-10-LABX-63). The work of BDW is supported by the
Simons Foundation.
\end{acknowledgements}

\bibliography{baylens}

\end{document}